# Computing Tails of Compound Distributions Using Direct Numerical Integration

**Xiaolin Luo**

CSIRO Mathematical and Information Sciences, Sydney, Locked bag 17, North Ryde, NSW, 1670, Australia. e-mail: Xiaolin.Luo@csiro.au

**Pavel V. Shevchenko**

CSIRO Mathematical and Information Sciences, Sydney, Locked bag 17, North Ryde, NSW, 1670, Australia. e-mail: Pavel.Shevchenko@csiro.au

**Abstract**
An efficient adaptive direct numerical integration (DNI) algorithm is developed for computing high quantiles and conditional Value at Risk (CVaR) of compound distributions using characteristic functions. A key innovation of the numerical scheme is an effective tail integration approximation that reduces the truncation errors significantly with little extra effort. High precision results of the 0.999 quantile and CVaR were obtained for compound losses with heavy tails and a very wide range of loss frequencies using the DNI, Fast Fourier Transform (FFT) and Monte Carlo (MC) methods. These results, particularly relevant to operational risk modelling, can serve as benchmarks for comparing different numerical methods. We found that the adaptive DNI can achieve high accuracy with relatively coarse grids. It is much faster than MC and competitive with FFT in computing high quantiles and CVaR of compound distributions in the case of moderate to high frequencies and heavy tails.

**Keywords**: characteristic function, compound distribution, truncation error, FFT, Monte Carlo

## 1. Introduction

The method of characteristic functions (CF) for computing probability distributions is a powerful tool in mathematical finance. In particular, it is used for calculating aggregate loss distributions in insurance, operational risk (OpRisk) and credit risk. It is also used for option price calculations in some models (e.g. exponentially affine models). Typically, the frequency-severity compound distributions can not be found in closed form but can be conveniently expressed through the inverse transform of the CFs. Similarly, in option pricing the expectations in the case of exponentially affine models can be more readily obtained using the CF method. In this paper we concentrate on computing high quantiles and conditional Value at Risk (CVaR) of compound distributions.

**Model**
Consider the compound random variable (rv), e.g. compound loss:

$$Z = \sum_{i=0}^{K} X_i, \qquad (1)$$

where $K$ is the number of events (frequency) over time period $T$ modelled as a discrete rv from a probability density function $p(k) = \Pr(K = k)$, $X_0 = 0$ and $X_i, i \geq 1$ are the severities of the events modelled as independent and identically distributed rvs from a continuous distribution function (df) $F(x)$ whose probability density function (pdf) will be denoted as $f(x)$. Note that there is a finite probability of no loss occurring over $T$ if $K = 0$ is allowed, i.e. $\Pr(Z = 0) = \Pr(K = 0)$. Here, we assume that the severities and frequency of the events are independent. Without loss of generality we set $T = 1$.

The CF of the severity density $f(x)$ is defined as

$$\varphi(t) = \int_{-\infty}^{\infty} f(x) e^{itx} dx, \qquad (2)$$

where $i = \sqrt{-1}$ is a unit imaginary number. Also, the probability generating function (pgf) of a discrete frequency pdf $p(k) = \Pr(K = k)$ is

$$\psi(s) = \sum_{k=0}^{\infty} s^k p(k). \qquad (3)$$

Then the CF of the compound loss $Z$ in model (1), denoted by $\chi(t)$, can be expressed through the pgf of the frequency distribution and CF of the severity distribution as

$$\chi(t) = \sum_{k=0}^{\infty} [\varphi(t)]^k p(k) = \psi(\varphi(t)). \qquad (4)$$

Hereafter, we consider the case of nonnegative severities, i.e. $f(x) = 0$ if $x < 0$. In this case the pdf and df of $Z$ can be calculated via the inverse Fourier transform as

$$h(z) = \frac{2}{\pi} \int_0^\infty \text{Re}[\chi(t)] \cos(tz) dt, \quad z \geq 0 \qquad (5)$$

and

$$H(z) = \frac{2}{\pi} \int_0^\infty \text{Re}[\chi(t)] \frac{\sin(tz)}{t} dt, \quad z \geq 0 \qquad (6)$$

respectively, see Appendix for proof. Hereafter, calculation of the df for $Z$ using (6) is referred to as *direct numerical integration* (DNI). The focus of this paper is the calculation of high quantiles of $H(z)$, i.e. $Q_q = H^{-1}(q)$, where $H^{-1}(q)$ is the inverse df and $q$ is a high quantile level, such as the 0.999 quantile required in OpRisk capital calculations, see BIS (2006).

**Numerical methods**

Typically, a Monte Carlo (MC) simulation method can be used to compute the distribution of $Z$. Although MC simulation is straightforward and robust, it is slow to get accurate results. High precision results are especially important for sensitivity studies, where the first or even the second order derivatives are involved. Fast Fourier Transform (FFT) and Panjer recursion (see Panjer (1981), Panjer and Willmot (1986)) are the other two popular alternatives for computing compound distribution. Both have a long history, but their applications to computing very high quantiles of the compound dfs with high frequencies and heavy tails are only recent developments and various pitfalls still exist. High precision results for these cases are not readily available in the literature, despite an increasing number of publications devoted to this area. For example, to deal with the 0.999 quantile in the case of high frequency and heavy tail dfs, the truncation point in FFT has to be high enough to exceed the 0.999 quantile, which conflicts with the requirement of fine grids for good accuracy, given limited computing resources.

A commonly recognised pitfall of FFT is the aliasing error in evaluating compound losses, which was recently studied in great detail in Embrechts and Frei (2008) and Schaller and Temnov (2008). These researchers applied the tilting or exponential windowing procedure to reduce this error. Only results for very moderate frequencies (e.g. $K$ is distributed from Poisson with intensity $\lambda < 50$) were shown in these studies. Hereafter, $\lambda$ is used to denote the intensity of Poisson distributed frequencies. In Schaller and Temnov (2008), a balance equation is derived to calculate the optimal value of the tilting parameter prior to the quantile calculation. This balance equation attempts to keep the balance between the aliasing error and the error caused by the interplay between exponential windowing and limited machine accuracy, minimizing the difference between the aliasing and numerical precision errors. As demonstrated, this optimal tilting is very effective in reducing the aliasing errors, in many cases obtaining a precision restricted by discretization error only. For heavy tail and high frequency, however, tilting alone may not be sufficient for high accuracy, due to the conflict requirements of fine grids for reducing discretization error and long integration domain for reducing truncation error. The exponential tilting technique for reducing aliasing error under the context of calculating compound distribution was first investigated by Grubel and Hermesmeier (1999).

The Panjer recursion has often been compared with FFT, and it is accepted that the former is slower if the grid size is large, see Bühlmann (1984); Grubel and Hermesmeier (1999); Embrechts and Frei (2008). In a recent paper, Peters, Johansen and Doucet (2007) utilize Panjer recursions, importance sampling and trans-dimensional Markov chain Monte Carlo to achieve a higher efficiency than the standard Monte Carlo method. Again, only results for Poisson distributed frequencies with small intensity parameter $\lambda = 2$ were presented.

Much work has been done in the last few decades in the general area of inverting CFs numerically, for example the works by Bohman (1975); Seal (1977); Abate and Whitt (1992, 1995); Heckman and Meyers (1983); Bühlmann (1984); Shephard (1991); Waller, Turnbull and Hardin (1995); and Den Iseger (2006), just to mention a few. These papers address various issues such as: singularity at the origin; treatment of long tails in the infinite integration; choices of quadrature rules covering different objectives with different distribution functions. We believe that no single approach is superior to all others under all circumstances. Craddock, Heath and Platen (2000) gave an extensive survey of numerical techniques for inverting CFs in the context of derivative pricing. They concluded that each of the many existing techniques has particular strengths and weaknesses, and no method works equally well for all classes of problems. For instance, there are special requirements in computing the 0.999 quantile of the aggregate loss distribution from a frequency-severity compound distribution. The accuracy demanded is high and at the same time the numerical inversion could be very time consuming due to rapid oscillations and slow decay in the CF, especially for the cases with large variance severities and high event frequencies. Methods that work satisfactorily for integrating analytically expressed functions could be too slow when applied to the CF of compound distributions which are themselves obtained numerically through semi-infinite integrations.

The specific objectives dictate the choice of the method. A tailor-made numerical algorithm for a specific task with a specific requirement on accuracy and efficiency is perhaps the best approach. Here, our specific goal is to develop and implement such an algorithm to calculate very high quantiles and the corresponding conditional Value at Risk (CVaR) for compound distributions. This is of high importance for OpRisk measurement. To our best knowledge we are not aware of any comprehensive, accuracy proven numerical results of the 0.999 quantiles and CVaR for very high event frequencies (e.g. $\lambda$ of the order of $10^5$) and heavy tailed severity distributions. This work was motivated partly by the lack of such comprehensive data in public domains, and partly by our recent need to compute high quantiles of OpRisk loss distribution in a systematic study on the impact of data truncation and parameter uncertainty (Luo, Shevchenko and Donnelly (2007)).

Many numerical inverting methods, including FFT, have two common features – they all have truncation or cut-off errors and use uniform grids in the numerical integration. That is, the semi-infinite integration is approximated by integration with a finite length, and the finite domain is uniformly subdivided. Usually this is fine because the cut-off error can always be reduced by extending the finite domain at the expense of computing time. However, when the CF itself is a semi-infinite integration and it oscillates and decays very slowly, then a sufficiently fine uniform partition covering a very long tail becomes computationally expensive. This is also the case where FFT could reach its limitation due to a very large grid size required for high quantiles of compound distributions with high frequency and heavy tail.

The method we propose here differs in both of the above mentioned common features. We use adaptive partition instead of a uniform one for the finite-domain approximation and we explicitly integrate the tail to infinity using a piecewise linear approximation. The adaptive partitioning reduces the total number of points required for evaluating the CF. The treatment of the tail reduces the cut-off error significantly to a much smaller discrete error, thus increasing overall accuracy of the inversion, while requiring almost no extra computing effort. This tail treatment allows a much shorter finite integration domain than it would be required otherwise, thus saving computer time. From another side, for the same length of finite-domain integration, the inclusion of the tail improves the accuracy significantly without much extra effort. The adaptive partitioning can also be considered as improving both accuracy and efficiency, because one is able to concentrate more quadrature points at where they are needed most.

**Remarks on high frequencies**

The mean of frequencies dealt with in this study ranges from 0.1 to $10^6$, which is more than sufficient to cover the possible high frequencies encountered in operational risk practice. The cases of Poisson frequencies with $\lambda \geq 10^5$ are not typically encountered in practice, but these calculations were performed mainly for stress testing our algorithm and for completeness. We would like to make the following remarks regarding high frequencies in OpRisk:

1. In practice, some OpRisks may have high frequencies, depending on granularity of data analysis. For example, results of the 2004 Loss Data Collection Exercise (LDCE), see the report of Federal Reserve System (2005), show at least four banks with average number of losses (*per year* for each bank, from 1999 to 2004) exceeding $68,000$. The actual frequency is higher than the observed, because losses are reported above a certain threshold and incomplete.
2. To estimate the true intensity from the observed loss events reported above a level *L*, a common practice is to fit the severity distribution $F(x)$ first, then the Poisson intensity is estimated as $\hat{\lambda} = \overline{K}_{obs}/(1-F(L))$, where $\overline{K}_{obs}$ is the observed average annual number of loss events. See e.g. Luo, Shevchenko and Donnelly (2007). Sometimes the estimated intensity is unrealistically high, e.g. $\hat{\lambda} > 10^6$. In this case Dutta and Perry (2007) suggested adjusting the threshold to obtain a more realistic estimate. To provide guidance and to study the impact on such an adjustment, the ability to compute high quantiles of the compound distribution at very high frequencies is useful. The total impact of those high frequency small losses can be ignored sometimes, but not always.
3. The merit of the DNI scheme does not mainly rely on very high frequencies. The DNI numerical scheme becomes competitive with FFT at least at $\lambda \geq 1000$, especially for CVaR for a given threshold. For large banks it is not uncommon to encounter $\lambda \sim 10,000$ for some business lines. The algorithm to deal with a wide range of frequencies with high accuracy is also helpful for sensitivity studies, model comparison and quantifying parameter uncertainty.

**Paper structure**

The paper is organized as follows. In Section 2, we describe the specific models for compound dfs and discuss general issues encountered in the CF inversion. Section 3 discusses numerical algorithm in details, in particular the adaptive spacing and the tail integration. A few examples are shown to demonstrate the effectiveness of the tail integration. Section 4 presents results of the 0.999 quantiles for compound dfs in the case of some frequency and severity dfs. Then results for the CVaR above a given threshold for Poisson-lognormal and negative binomial-lognormal compound dfs are presented. The DNI results for the 0.999 quantiles and CVaR for mean frequency ranging from 0.1 to $10^6$ are compared with the FFT and MC results, in terms of the accuracy and speed. Whenever possible, $10^8$ MC simulations were carried out for the accuracy comparison. Concluding remarks are given in Section 5.

## 2. Model distributions and general numerical issues

The numerical algorithm presented in this paper should work for a variety of frequency and severity distributions. For illustrative purposes we assume that the frequency $K$ is modelled either by the Poisson or the negative binomial distribution, and the severity is modelled by either lognormal or generalized Pareto distribution (GPD). These distributions cover the cases most relevant to OpRisk management practice. The various densities are given as follows.

- The Poisson density, $Poisson(\lambda)$:

$$\Pr(K = k \mid \lambda) = \frac{\lambda^k}{k!} \exp(-\lambda), \quad \lambda > 0, k = 0,1,\dots \quad , \tag{7}$$

with $\text{mean}(K) = \text{var}(K) = \lambda$. Note that for Poisson frequency, it can be shown from (6) that $H(0) = e^{-\lambda}$, reflecting that there is a finite probability of zero loss.

- The negative binomial density, $NegBinomial(p,m)$:

$$\Pr(K = k \mid p,m) = \binom{k+m-1}{k}(1-p)^k p^m, 1 \geq p \geq 0, m \geq 1, \tag{8}$$

with $\text{mean}(K) = m(1-p)/p$, $\text{var}(K) = m(1-p)/p^2$. In this case $H(0) = p^m$.

- The lognormal density, $Lognormal(\mu,\sigma)$:

$$f(x \mid \mu,\sigma) = \frac{1}{x\sqrt{2\pi\sigma^2}} \exp\left(-\frac{(\ln x - \mu)^2}{2\sigma^2}\right), \quad \sigma > 0, \ 0 < x < \infty \ , \tag{9}$$

- The generalized Pareto density, $GPD(\xi,\beta)$:

$$f(x \mid \xi,\beta) = \frac{1}{\beta}\left(1 + \frac{\xi x}{\beta}\right)^{-1-1/\xi}. \tag{10}$$

For $GPD$ we consider $\xi > 0, 0 \leq x < \infty$. In this case some moments do not exist, e.g. variance and higher moments do not exist if $\xi \geq 0.5$.

The CF (4) of compound loss (1) in the case of $K$ being distributed as $Poisson(\lambda)$ is

$$\chi(t) = \sum_{k=0}^{\infty} [\varphi(t)]^k \frac{e^{-\lambda} \lambda^k}{k!} = \exp[\lambda \varphi(t) - \lambda] \qquad (11)$$

and in the case of *NegBinomial(p,m)* frequency, it is

$$\chi(t) = \sum_{k=0}^{\infty} [\varphi(t)]^k \binom{k+m-1}{k} (1-p)^k p^m = \left( \frac{p}{1-(1-p)\varphi(t)} \right)^m. \qquad (12)$$

The explicit expression of $\text{Re}[\chi(t)]$ for *Poisson(λ)* is

$$\text{Re}[\chi(t)] = e^{-\lambda} \exp(\lambda \text{Re}[\varphi(t)]) \times \cos(\lambda \text{Im}[\varphi(t)]). \qquad (13)$$

For the *NegBinomial(p,m)* case, $\text{Re}[\chi(t)]$ is easily obtained through complex variable functions in the relevant computer language.

The task of the CF inversion is analytically straightforward, but numerically difficult in terms of achieving high accuracy and computational efficiency *simultaneously*. The computation of compound df through the CF involves two steps: computing the CF (Fourier transform of pdf, referred to as the forward integration) and inverting it (referred to as the inverse integration). The first step, the integration of (2), is relatively easier because the severity pdf to be transformed typically has closed form expression, and is well-behaved having a single mode. This step can be done more or less routinely and many existing algorithms, including the ones commonly available in many software packages, can be employed. Then the CF of compound loss is calculated using (4).

However, the second step, the integration of (6), is much more challenging. To start with, each single integrand point in the inversion step is obtained *numerically* through the first step, a semi-infinite integration. Efficiency which is not a big issue in the first step will now become critical. The total number of forward integrations required by the inversion is usually quite large, because in this case the CF could be highly oscillatory due to high frequency and it may decay very slowly due to heavy tails, as will be shown later. A fine resolution over a large region approximating a semi-infinite domain is computationally intensive, when each integrand value itself is a semi-infinite integration. The same resolution and cut-off strategy which worked well in the first step may not work in the second step - the failure could be either insufficient accuracy or too long computing time or both. Below we address the accuracy and efficiency issues discussed above.

## 3. Adaptive direct numerical integration scheme

In principle, if computing cost is ignored one can almost always obtain an accurate CF inversion by subdividing the semi-infinite integration domains into sub-regions as small as required, using numerical quadrature with the order of accuracy as high as required, and taking the finite domain as large as required to reduce truncation error. It is the dual requirement of high accuracy and efficiency that makes the task a challenge.

A typical accuracy requirement on the df evaluation can be shown with a simple example of the lognormal distribution with $\mu = 0$ and $\sigma = 2$. In this case, the "exact" 0.999 quantile $Q_{0.999} = 483.2164...$. However, at $q = 0.99902$, the quantile becomes $Q_q = 489.045...$, i.e. a mere 0.002% change in the df value causes more than 1% change in the quantile value, which is an amplification of error by 500 times in percentage terms. In other words, to limit the error for the 0.999 quantile within 1% requires the calculation of df to be accurate to the fifth digit or the relative error less than $0.002\%$.

Formally, the error propagation from the df level to the quantile value can be estimated by the relation between the pdf, $f(x)$, and its df, $F(x)$, $dF/dx = f(x)$. In the above example, $x = 483.2164...$, $1/(dF/dx) = 1/f(x) = \sigma x\sqrt{2\pi} \exp[0.5(\ln x/\sigma)^2] \approx 287023$. That is, in absolute terms, an error in the df estimation will be amplified by 287023 times in the error for the corresponding 0.999 quantile. In the case of a compound df, the requirement for accuracy in df could be even higher than demonstrated here, because $1/f(x)$ could be larger at $x = Q_{0.999}$ for the compound case. In fact, for compound df with high frequency and heavy tails, we often observed that df correct to the fifth digit is not accurate enough for accurate estimation of the 0.999 quantiles. Below we describe DNI algorithm for compound dfs.

### *3.1. The forward integration*

The building blocks are the real and imaginary parts of the CF for a severity distribution. In the case of non-negative severities considered in this paper, the required forward integrations are given by

$$\text{Re}[\varphi(t)] = \int_0^\infty f(x)\cos(tx)dx, \tag{14}$$

$$\text{Im}[\varphi(t)] = \int_0^\infty f(x)\sin(tx)dx. \tag{15}$$

The severity pdf, given by (9) or (10), has a single mode, which means that the oscillatory nature of the integrand only comes from the sin( ) or cos( ) functions. This well-behaved weighted oscillatory integrand can be effectively dealt with by the modified Clenshaw-Curtis integration method, see Clenshaw and Curtis (1960); Piessens, Doncker-Kapenga, Überhuber and Kahaner (1983). In this method the oscillatory part of the integrand is transferred to a weight function, the non-oscillatory part is replaced by its expansion in terms of a finite number of Chebyshev polynomials, and the modified Chebyshev moments are calculated. If the oscillation is slow when the argument $t$ of the CF in (14) and (15) is small, the standard Guass-Legendre and Kronrod quadrature formulae, Kronrod (1965); Golub and Welsh (1969); Szegö (1975), are more effective. We have used *IMSL (Numerical Libraries, Fortran Version 3.0)* functions utilizing the Guass-Kronrod quadrature to perform the above forward integrations (14) and (15).

Let $\delta_m^G$ denote the error bound for the $m$-order Gauss quadrature and $\delta_{2m+1}^{GK}$ be the error bound for the corresponding Guass-Kronrod quadrature. Brass and Förster (1987) proved that $\delta_{2m+1}^{GK}/\delta_m^G \leq const \times \sqrt[4]{m}(1/3.493)^m$. Because $\delta_{2m+1}^{GK}$ is smaller than $\delta_m^G$ by at least an order of magnitude, the difference between Gauss-Kronrod and Gauss quadrature serves as a good estimate for $\delta_m^G$. The *IMSL* functions use this estimate to achieve an overall error bound below the user specified tolerance. In general, for the forward integrations (14-15), double precision accuracy can be routinely achieved. The accuracy of the CF calculation can be checked by applying the method to simple severity dfs (without compounding), where closed form or double precision df is available.

## 3.2. The inverse integration

Changing variable $x = tz$, (6) can be rewritten as

$$H(z) = \frac{2}{\pi} \int_0^\infty \frac{\text{Re}[\chi(x/z)]}{x} \sin(x) dx, \quad (16)$$

where $\chi(t)$ depends on $\text{Re}[\varphi(t)]$ and $\text{Im}[\varphi(t)]$ calculated from the forward integrations (14) and (15) as discussed above for any required argument $t$. In the case of *Poisson* and *NegBinomial*, see (13) and (12). Obviously, the above integration is more difficult than the forward integration, as there are two oscillatory components represented by $\sin(x)$ and another part in $\text{Re}[\chi(x/z)]$. For example, for Poisson frequency $\text{Re}[\chi(t)] = e^{\lambda(\text{Re}[\varphi(t)]-1)} \cos(\lambda \text{Im}[\varphi(t)])$.

### 3.2.1. Adaptive partition

It is convenient to treat $\sin(x)$ as the principal oscillatory factor and the other part as secondary. Define

$$G(x) = \frac{2}{\pi} \frac{\text{Re}[\chi(x/z)]}{x}, \quad (17)$$

where the explicit dependence of $G(x)$ on $z$ is dropped for notational convenience, and rewrite (16) as

$$H(z) = \int_0^\infty G(x) \sin(x) dx. \quad (18)$$

Typically, given $z$, $\text{Re}[\chi(x/z)]$ decays fast initially and then approaches zero slowly as $x$ approaches infinity. For example, see Figure 1, which shows plot of $\text{Re}[\chi(x/z)]$ as a function of $x$ in the case of *Poisson*($10^5$)-*Lognormal*(0,2) compound distribution with the value of $z$ corresponding to $H(z) = 0.999$.

Although the oscillation frequency of $\text{Re}[\chi]$ increases with $\lambda$, this increase is much slower than a linear increase. In fact, at $\lambda = 10^5$ (see Figure 1) the oscillation frequency of $\text{Re}[\chi]$ is still smaller than that of $\sin(x)$, the principle oscillator. This can be quantified by $\omega$, the relative oscillation frequency of $\text{Re}[\chi]$ with respect to $\sin(x)$, defined as

$$\omega(x,z) = \lambda \frac{\partial \text{Im}[\varphi(x/z)]}{\partial x}, \quad (19)$$

where $\omega < 1$ indicates that the local oscillation frequency is smaller than that of $\sin(x)$. Figure 2 shows a plot of $\omega$ as a function of $x$. It shows that not only $\omega$ is less than one in this case, but also that it appears to decay linearly as $x$ increases, justifying treatment of $\text{Re}[\chi]$ as the secondary oscillator. Nevertheless, our numerical algorithm is designed to handle the situation when the frequency of $\text{Re}[\chi]$ is higher than $\sin(x)$.

We could apply some standard general purpose integration schemes, such as those available in *IMSL*, to properly sub-divided sections. Typically these schemes are globally adaptive, for example, the *IMSL* subroutine *QDAG* subdivides a given interval and uses the ($2m+1$)-point Gauss-Kronrod rule to estimate the integral over each subinterval. The error for each subinterval is estimated by comparison with the *m*-point Gauss quadrature rule. The subinterval with the largest estimated error is then bisected and the same procedure is applied to both halves. The bisection process is continued until the error criterion is satisfied, or the subintervals become too small, or the maximum number of subintervals allowed is reached. This is not ideal for our case. Firstly, it does not address irregular oscillation specifically, thus the general purpose adaptive procedure may not be the most efficient. Secondly, the iterative process can lead to excessive number of integrand function evaluations, which is very time consuming when the integrand itself has to be numerically obtained through semi-infinite integrations. We have implemented the following adaptive scheme instead.

The integral in (18) is now divided into intervals with an equal length of $\pi$ (referred to as $\pi$-cycle):

$$H(z) = \sum_{k=0}^{\infty} H_k, \quad H_k = \int_{k\pi}^{(k+1)\pi} G(x)\sin(x)dx. \tag{20}$$

Within each $\pi$-cycle, the secondary oscillation could be dominating for some early cycles, thus the $\pi$-cycle could in fact contain multiple cycles due to the "secondary" oscillation. So a further sub-division is warranted. One of the drawbacks with some algorithms like FFT is that it uses uniform grids across the entire domain for both forward and inverse transforms. Sub-dividing interval $(k\pi,(k+1)\pi)$ into $n_k$ segments of equal length of $\Delta_k = \pi/n_k$, (20) can be written as

$$H_k = \sum_{j=1}^{n_k} H_k^{(j)}, \quad H_k^{(j)} = \int_{k\pi+(j-1)\Delta_k}^{k\pi+j\Delta_k} G(x)\sin(x)dx. \tag{21}$$

For convenience denote $a_{k,j} = k\pi + (j-1)\Delta_k$ and $b_{k,j} = a_{k,j} + \Delta_k$. The above calculation will be most effective if the sub-division is made adaptive for each $\pi$-cycle according to the changing behaviour of $G(x)$. Assuming that for the first $\pi$-cycle ($k=0$) we have initial partition $n_0$, for the subsequent cycles we make $n_k$ adaptive by the following two simple rules:

Rule 1. Let $n_k$ be proportional to the number of $\pi$-cycles of the secondary oscillation – the number of oscillations in $G(x)$ within each principal $\pi$-cycle;
Rule 2. Let $n_k$ be proportional to the magnitude of the maximum gradient of $G(x)$ within each principal $\pi$-cycle.

The theoretical basis of the above two rules is simple: the *m*-point Gaussian quadrature for each partition makes the computed integral exact for all polynomials of degree $2m-1$ or less; and the above adaptive strategy directly limits the order of the polynomial function that can accurately represent the integrand in each subdivision. Rule 1 limits the number of oscillations in each partition, and Rule 2 ensures that the integrand does not change too rapidly in magnitude within each subdivision. A slower oscillation frequency and a smaller maximum gradient on a given interval correspond to a lower order polynomial. Rule 1 effectively counts the number of distinctive roots of the polynomial within each subdivision. Rule 2 makes the grid size $\Delta_k$ inversely proportional to the magnitude of the local maximum gradient. Of course, the precise degree of polynomials required is not known for a given precision, and the associated error will be discussed in the next section. For functional flexibility, we also let $n_k$ be responsive to user's input by allowing user to specify $n_0$ for the first cycle. For subsequent cycles, $n_k$ is adaptively updated according to the above two rules and proportional to $n_0$, therefore the sub-division is not only adaptive, but also under the user's control, thus fine or coarse grids can be specified by the user as desired. This feature is important for a robust numerical estimate of errors.

Application of Rule 1 and Rule 2 requires correct counting of secondary cycles and good approximation of the local gradient in $G(x)$. Both can only be achieved with a significant number of points at which $G(x)$ is computed within each cycle. Using the $m$-point Gaussian quadrature as described below, we have a total of $m \times n_k$ points for the estimates of secondary cycles and local gradient for each $\pi$-cycle.

### *3.2.2. Gaussian quadrature for each subdivision*

With a proper sub-division, even a simple trapezoidal rule can be applied to get a good approximation for integration over the sub-division $H_k^{(j)}$ in (21). However, higher order numerical quadrature can achieve higher accuracy for the same computing effort or it requires less computing effort for the same accuracy. The $m$-point Gaussian quadrature makes the computed integral exact for all polynomials of degree $2m-1$ or less. The efficiency of the Gaussian quadrature is much superior to the trapezoidal rule. For instance, integrating the function $\sin(3x)$ over the interval $(0, \pi)$, the 7-point Gaussian quadrature has a relative error less than $10^{-5}$, while the trapezoidal rule requires about 900 function evaluations (grid spacing $\delta x = \pi / 900$) to achieve a similar accuracy. The reduction of the number of integrand function evaluations is important for a fast integration of (20), because the integrand itself is a time consuming semi-infinite numerical integration.

The integration over the sub-division $(a_{k,j}, b_{k,j})$, by the $m$-point Gaussian quadrature rule, is then

$$H_k^{(j)} \approx \frac{\Delta_k}{2} \sum_{i=1}^{m} w_i G(x_{k,j}^i) \sin(x_{k,j}^i), \quad x_{k,j}^i = (\zeta_i \Delta_k + a_{k,j} + b_{k,j})/2, \qquad (22)$$

where $0 < w_i < 1$ and $-1 < \zeta_i < 1$ are the $i^{\text{th}}$ weight and the $i^{\text{th}}$ abscissa of the Gaussian quadrature respectively, and $m$ is the order of the Gaussian quadrature. Throughout this work we have used the 7-point Gaussian quadrature ($m = 7$), which computes all polynomials of degree 13 or less exactly. Because our sub-division strategy, as discussed above, ensures the integrand within the sub-division is monotonic and does not change too rapidly, certainly a polynomial of degree 13 or less would present a virtually "exact" presentation of the integrand within each sub-division.

The error of the $m$-point Gaussian quadrature rule can be accurately estimated if the $2m$ order derivative of the integrand can be computed (Kahaner, Moler and Nash (1989); Stoer and Bulirsch (2002)). For an integrand $g(x)$ which has $2m$ continuous derivatives, the error is

$$\delta I = \left| \int_a^b g(x)dx - \frac{\Delta}{2} \sum_{i=1}^m w_i g\big((a+b+\zeta_i\Delta)/2\big) \right| = \frac{\Delta^{2m+1}(m!)^4}{(2m+1)[(2m)!]^3} \left| g^{(2m)}(\eta) \right|, \quad a < \eta < b, \quad (23)$$

where $g^{(2m)}$ is the $2m$-th derivative and $\Delta = b - a$. For $m = 7$, we have

$$\delta I = \frac{\Delta^{15} \times (7!)^4}{15 \times (14!)^3} \left| g^{(2m)}(\eta) \right| < 10^{-19} \times C \times \Delta^{15}, \quad (24)$$

where $C = \left| g^{(14)}(\eta) \right|$. In the case of $g(x) = e^{-\alpha x}$ (a fast changing function near $x = 0$ for large $\alpha$, not very different from our case of compound CF) or $g(x) = \sin(\alpha x)$ (a rapid oscillating function for large $\alpha$, also a feature of compound CF), we have $|g^{(14)}(\eta)| < \alpha^{14}$. Thus $|\delta I| < 10^{-19}(\alpha \Delta)^{14}\Delta$, and by letting $\Delta \leq \min(1, 1/\alpha)$ we can readily achieve double precision accuracy. In general, double precision accuracy is assured by letting $\Delta \leq \min(1, C^{-1/15})$. Stoer and Bulirsch (2002) commented that this error estimate is inconvenient in practice, since in general it is difficult to estimate the order $2m$ derivative and in addition the actual error may be much less than a bound established by the derivative. An accepted good practice is to use two numerical evaluations of different grid sizes and estimate the error as the difference between the two results. For example, if we use $\delta$ in the first evaluation to get result $I_1$ and use $\delta/2$ in the second evaluation to get result $I_2$, the error estimate for the first evaluation can be approximated simply by $\delta I_1 \approx |I_2 - I_1|$. This is because $\delta I_2$ is of the order of $0.5^{15} \delta I_1$ according to (24), which is several order of magnitude smaller than $\delta I_1$. Since our partition strategy described in the previous section allows the grid size of all the subsequent sub-divisions to be proportional to the initial grid size specified by the user, we can have a convenient estimate of the discrete error resulting from quadrature (22). Equivalently, different orders of quadrature can be used for the same grid to estimate error.

For $2N$ $\pi$-cycles, i.e. setting truncation point at $2\pi N$, where $N$ denotes the total number of full cycles before truncation point, the entire integration (20) becomes

$$H(z) \approx \sum_{k=0}^{2N-1} \frac{\Delta_k}{2} \left[ \sum_{j=1}^{n_k} \sum_{i=1}^m w_i G(x_{k,j}^i) \sin(x_{k,j}^i) \right]. \quad (25)$$

### 3.2.3. Tail integration

The truncation error of using (25) is simply

$$H_T = \int_{2N\pi}^{\infty} G(x)\sin(x)dx. \quad (26)$$

For higher accuracy, instead of increasing truncation length at the cost of computing time, we propose to compute the tail integration $H_T$ explicitly by a very economical but effective simplification, taking advantage of $G(x)$ varying slowly (as imaginary part goes to zero) and approaching zero as $x \to \infty$. Integrating (26) by parts, we have

$$\int_{2N\pi}^{\infty} G(x)\sin(x)dx = G(2N\pi) - \int_{2N\pi}^{\infty} G^{(1)}(x)\cos(x)dx, \tag{27}$$

where $G^{(1)}(x) = \partial G(x)/\partial x$. If we assume $G(x)$ is linear within each $\pi$-cycle in the tail, then the integration $\int_{2N\pi}^{\infty} G^{(1)}(x)\cos(x)dx$ vanishes, because within each $\pi$-cycle $G^{(1)}(x)$ is constant from the piecewise linear assumption and $\int_{k\pi}^{(k+1)\pi} \cos(x)dx = 0$ for any integer $k$, and we assume that $G^{(1)}(\infty) \to 0$. So under the piecewise linear assumption, (27) becomes

$$H_T = \int_{2N\pi}^{\infty} G(x)\sin(x)dx \approx G(2N\pi). \tag{28}$$

Equation (28) gives a simple formula to compute the tail integration. This elegant result means that we only need to evaluate the integrand (17) at one single point $x = 2\pi N$ (the truncation point) for the entire tail integration, replacing the truncation error with a much smaller round-off error. As will be demonstrated later, this one-point formula for the potentially demanding tail integration is remarkably effective in reducing the truncations errors caused by ignoring $H_T$. Continuing with integration by parts in (27) and assuming $G^{(1)}(x) \to 0$ at infinity, we obtain

$$\int_{2N\pi}^{\infty} G(x)\sin(x)dx = G(2N\pi) - \int_{2N\pi}^{\infty} G^{(2)}(x)\sin(x)dx, \tag{29}$$

where $G^{(2)}(x) = \partial^2 G(x)/\partial x^2$. Equation (29), as well as (27), is exact – no approximation is involved. The recursive pattern in (29) is evident. If we assume the second derivative $G^{(2)}(x)$ is piecewise linear in the tail, also an asymptotically valid assumption because $G^{(2)}(x)$ must also go to zero at infinity, then (26) becomes

$$H_T = \int_{2N\pi}^{\infty} G(x)\sin(x)dx \approx G(2N\pi) - G^{(2)}(2N\pi). \tag{30}$$

With the additional correction term, (30) is more accurate than (28). In general, without making any approximation, from the recursive pattern of (29) we arrive at the following expression for the error associated with formula (28)

$$\varepsilon_T = \int_{2N\pi}^{\infty} G(x)\sin(x)dx - G(2N\pi) = \sum_{k=1}^{\infty} (-1)^k G^{(2k)}(2N\pi), \tag{31}$$

where $G^{(2k)}(2N\pi)$ is the $2k$ order derivative of $G(x)$ at the truncation point. The absolute error of tail integration $|\varepsilon_T|$ is denoted as $\delta_T$. As will be shown later with examples, typically the first few terms from (31) are sufficiently accurate. Note that (31) is convergent if $G^{(2k)}(x) \to 0$ when $k \to \infty$. This is not the case for some functions such as $e^{-x}$ (however, (31) still goes to zero with increasing $N$). Finally, the entire integration is approximated by

$$H(z) = \sum_{k=0}^{\infty} H_k \approx \sum_{k=0}^{2N-1} \frac{\Delta_k}{2} \left[ \sum_{j=1}^{n_k} \sum_{i=1}^{m} w_i G(x_{k,j}^i)\sin(x_{k,j}^i) \right] + G(2N\pi). \tag{32}$$

**Remarks**. The total error of (32) is bounded by the sum of three errors: the error of the Gaussian quadrature; the error of the tail integration; and the error propagated from the error of the forward integration, which will be discussed later. Note that there is no truncation error in (32) and thus all errors are discretization error in nature. In theory, the Gaussian quadrature error can be estimated by (24) and the tail integration error by (31). In practice, however, derivatives of compound distribution CF at the truncation point can only be evaluated numerically. As mentioned earlier, a common approach to estimate error is to reduce the gird size or equivalently use higher order quardratures for the same grid. The highly efficient Gauss–Kronrod rules are based on such an approach. For the tail integration, error estimation involves increasing the truncation length. The bound of the total error associated with (32) will be discussed in more detail in Section 3.3.

The assumption of piecewise linearity, although reasonable for $G(x)$ at large $x$, may appear to be rather crude for a high precision computation. However, we recall that we are only trying to reduce the already small truncation error $H_T$ and any reasonable approximation in $H_T$ could lead to significant improvement in the overall accuracy of integration. For example, suppose a relative error of 1% due to ignoring truncation using (25) and 10% error in evaluating the tail integration using the very simple formula (28). The overall accuracy with this tail integration added is now improved from 1% to 0.1% (1% times 10%). This improvement by an order of magnitude is achieved by simply evaluating the integrand at the truncation point. The assumption of a piecewise linearity applies to a broad range of functions, not just necessarily related to CFs, thus the special tail integration approximation can have a much wider application. Note, piecewise linear assumption does not even require monotonicity - $G(x)$ can be oscillating, as long as its frequency is relatively small compared with the principal cycles, as demonstrated in one of the examples below.

If the oscillating factor is $\cos(x)$ instead of $\sin(x)$, we can still derive a one-point formula similar to (28) by starting the tail integration at $(2N-1/2)\pi$ instead of $2N\pi$. In this case, the tail integration is

$$\int_{(2N-1/2)\pi}^{\infty} G(x)\cos(x)dx \approx G((2N-1/2)\pi).$$

Also, the tail integration approximation can be applied to the left tail (integrating from $-\infty$ to $-2N\pi$) as well, if such integration is required.

### 3.2.4. Examples of tail integration

The effectiveness of the above tail integration approximation is now demonstrated in a few examples. Introduce the following notations

$$I_E = \int_0^{\infty} G(x)\sin(x)dx,$$

$$\tilde{I}(2N\pi) = \int_0^{2\pi N} G(x)\sin(x)dx,$$

$$I_T(2N\pi) = \int_{2\pi N}^{\infty} G(x)\sin(x)dx.$$

In all the following examples the exact semi-infinite integration $I_E$ is known in closed form, and its truncated counterpart $\tilde{I}(2N\pi)$ is either known in closed form or can be computed accurately. The exact tail integration $I_T(2N\pi)$ can be computed from $I_T(2N\pi) = I_E - \tilde{I}(2N\pi)$. We compare $\tilde{I}(2N\pi) + G(2N\pi)$ with $\tilde{I}(2N\pi)$ and compare both of them with the exact semi-infinite integration $I_E$. The error reduction can be quantified by comparing the "magic" point value given by formula (28) with the exact tail integration $I_T(2N\pi)$. The error of using (28), $\varepsilon_T = I_E - [\tilde{I}(2N\pi) + G(2N\pi)]$, is given by (31).

***Example 1:*** $G(x) = e^{-\alpha x}$, $(\alpha > 0)$.
In this example, the closed form results are

$$I_E = \int_0^\infty e^{-\alpha x} \sin(x) dx = \frac{1}{1+\alpha^2}, \quad \tilde{I}(2N\pi) = \int_0^{2N\pi} e^{-\alpha x} \sin(x) dx = \frac{1-e^{-2\alpha N\pi}}{1+\alpha^2}.$$

Figure 3a compares the "magic" point value $G(2N\pi)$ representing simplified tail integration with the exact tail integration $I_T(2N\pi) = I_E - \tilde{I}(2N\pi)$ as functions of parameter $\alpha$ for $N=5$, i.e. the truncated lengths $l_T = 2N\pi = 10\pi$. The figure shows that a simple formula (28) matches the exact semi-infinite tail integration surprisingly well for the entire range of parameter $\alpha$. Figure 3b shows the same comparison at an even shorter truncated length of $4\pi$ ($N=2$). The error of using (28) is $\delta_T = \left|\alpha^2 \exp(-2\alpha\pi N)/(1+\alpha^2)\right|$. If $\alpha$ is large, the function $G(x) = e^{-\alpha x}$ is "short tailed" and it goes to zero very fast. The absolute error $\delta_T$ is very small even at $N=2$. The relative error (against the already very small tail integration), given by $\delta_T / \left|I_E - \tilde{I}(2N\pi)\right| = \alpha^2$, is actually large in this case. But this large relative error in the tail approximation does not affect the high accuracy of the approximation for the whole integration. What is important is the error of the tail integration relative to the whole integration value. Indeed, relative to the exact integration, the error of using (28) is $\delta_T / |I_E| = \alpha^2 \exp(-2\alpha\pi N)$, which is about $2.7 \times 10^{-53}$ at $N=2$.

For a small value of parameter $\alpha$, the truncation error will be large unless the truncated length is very long. For instance, with $\alpha = 0.01$ the truncation error (if ignore the tail integration) is more than 70% at $l_T = 10\pi$ ($N=5$, as the case in Figure 3a), and it is more than 88% at $l_T = 4\pi$ ($N=2$, as the case in Figure 3b). On the other hand, if we add the "magic" value from formula (28) to approximate the tail integration, the absolute error of the complete integration $\delta_T$ due to this approximation is less than 0.01%, and the relative error is $\delta_T = \alpha^2 = 0.01\%$ at both $l_T = 10\pi$ and $l_T = 4\pi$. In other words, by including this one-point value, the accuracy of integration has dramatically improved by several orders of magnitude at virtually no extra cost, compared with the truncated integration. For the truncated integration $\tilde{I}(2N\pi)$ to have similar accuracy as $\tilde{I}(4\pi) + G(4\pi)$, we need to extend the truncated length from $4\pi$ to $300\pi$ for this heavy tailed integrand.

***Example 2:*** $G(x) = 1/\sqrt{x}$.
This example has a heavier tail than the previous one. Here, we have closed form for $I_E$, but not for $\tilde{I}$ or $I_T$,

$$I_E = \int_0^\infty \frac{\sin(x)}{\sqrt{x}} dx = \sqrt{\frac{\pi}{2}}, \quad \tilde{I}(2N\pi) = \int_0^{2N\pi} \frac{\sin(x)}{\sqrt{x}} dx.$$

$\tilde{I}$ or $I_T$ can be accurately computed by a *IMSL* function using the modified Clenshaw-Curtis integration method (Clenshaw and Curtis (1960); Piessens, Doncker-Kapenga, Überhuber and Kahaner (1983)) as described in Section 3.1.

Figure 4a compares the "exact" tail integration $I_T(2N\pi) = \int_{2N\pi}^{\infty} \sin(x)/\sqrt{x}dx$ with our one-point value $G(2N\pi)$. Again the one-point approximation does an extremely good job. Even at the shortest truncation length of just $2\pi$ ($N=1$) the one-point approximation is very close to the exact semi-infinite tail integration. Applying the analytical error formula (31) to $G(x) = 1/\sqrt{x}$, we have

$$\varepsilon_T = \sum_{k=1}^{\infty} (-1)^k \frac{1 \times 3 \times ... \times (4k-1)}{2^{2k} x^{(4k+1)/2}}, \quad x = l_T = 2\pi N.$$

Taking the first three leading terms we get $\varepsilon_T \approx -0.007578995 + 0.001679809 - 0.001053114$ at $N=1$ and $\varepsilon_T \approx -2.39669 \times 10^{-5} + 5.31202 \times 10^{-8} - 3.33024 \times 10^{-10}$ at $N=10$. The relative error $\delta_T / I_E(2N\pi)$ is about 1% at $N=1$ and it is about 0.002% at $N=10$. Apparently, if the extra correction term $G^{(2)}(2N\pi)$ is included as in (30), the error $\delta_T$ reduces further by an order of magnitude at $N=1$ and by several orders of magnitude at $N=10$.

Figure 4b shows the truncated integration $\tilde{I}(2N\pi)$ and the truncated integration with the tail modification (28) added, i.e. $\tilde{I} + G(2N\pi)$, along with the correct value of the full integration $I_E = \sqrt{\pi/2}$. The contrast between results with and without the one-point tail approximation is striking. At the shortest truncation length of $2\pi$ ($N=1$), the relative error due to truncation for the truncated integration $(I_E - \tilde{I}(2N\pi))/I_E$ is more than 30%, but with the tail approximation added, the relative error $(I_E - \tilde{I}(2N\pi) - G(2N\pi))/I_E$ reduces to 0.5%. At $100\pi$, the largest truncation length shown in Figure 4b, the relative error due to truncation is still more than 4%, but after the "magic" point value is added the relative error reduces to less than $0.5 \times 10^{-6}$.

Another interesting way to look at these comparisons, which is relevant for integrating heavy tailed functions, is to consider the required truncation length for the truncated integration to achieve the same accuracy as the one with the "magic" value added. For the truncated integration $\tilde{I}(2N\pi)$ to achieve the same accuracy of $\tilde{I}(2\pi) + G(2\pi)$ (integration truncated at one-cycle plus the "magic point value), we need to extend the integration length to $7700\pi$. For $\tilde{I}(2N\pi)$ to achieve the same accuracy of $\tilde{I}(100\pi) + G(100\pi)$, the integration length has to be extended to more than $10^{12}\pi$! On the other hand, if we add the tail approximation $G(7700\pi)$ to $\tilde{I}(7700\pi)$, the relative error reduces from 0.5% to less than $10^{-11}$! This error reduction requires no extra computing, since $G(7700\pi)$ is simply a number given by $1/\sqrt{7700\pi}$.

*Example 3:* $G(x) = \cos(\alpha x)/x, \alpha < 1$.

We have remarked that the piecewise linear assumption does not require monotonicity, i.e. $G(x)$ can be oscillating, as long as its frequency is relatively small compared with the principal cycles. This example demonstrates this important point, as the function $G(x)$ encountered in compound CF is oscillating and its frequency approaches zero in the tail. In this example there is a closed form for $I_E$, but not for $\tilde{I}$ or $I_T$,

$$I_E = \int_0^{\infty} \frac{\cos(\alpha x)\sin(x)}{x}dx = \frac{\pi}{2}, \quad \tilde{I}(2N\pi) = \int_0^{2N\pi} \frac{\cos(\alpha x)\sin(x)}{x}dx, \quad \alpha < 1.$$

Figure 5a compares the "exact" tail integration $I_T(2N\pi)$ with the one-point approximation $G(2N\pi)$ for the case $\alpha = 0.2$. Again the one-point approximation performs surprisingly well, despite $G(x)$ itself is now an oscillating function, along with the principal cycles in $\sin(x)$. The piecewise linearity assumption is apparently still valid for relatively mild oscillating $G(x)$.

Figure 5b compares the truncated integration $\tilde{I}(2N\pi)$ against $\tilde{I} + G(2N\pi)$, along with the correct value of the full integration $I_E = \pi/2$. At truncation length $6\pi$ ($N = 3$), the shortest truncation length shown in Figures 5a and 5b, the relative error $\delta_T / I_E$ is less than 0.06% and it is less than 0.01% at $N = 50$. In comparison, the truncated integration without the end point correction has relative error of 2.7% and 0.2%, respectively for those two truncation lengths. Applying the analytical error formula (31) to $G(x) = \cos(\alpha x)/x$ and noting $\sin(\alpha x) = 0$ and $\cos(\alpha x) = 1$ with $\alpha = 0.2$ and $x = 2N\pi = 100\pi$, we obtain

$$\varepsilon_T \approx -\left(-\frac{\alpha^2}{x} + \frac{2}{x^3}\right) + \left(\frac{\alpha^4}{x} - \frac{12\alpha^2}{x^3} + \frac{24}{x^5}\right), \quad x = l_T = 100\pi,$$

where only the first two leading terms corresponding to the 2nd and 4th derivatives are included, leading to $\varepsilon_T \approx 0.0001273 + 5.07749 \times 10^{-6}$ at $N = 50$ that agrees with the actual error. Similar to the previous example, if we include the extra correction term $G^{(2)}(2N\pi)$, the error reduces further by two orders of magnitude at $N = 50$.

All these examples show dramatic reduction in truncation errors if tail integration formula (28) is employed, with virtually no extra cost. If the extra correction term $G^{(2)}(2N\pi)$ is included, i.e. using (30) instead of (28), the error is reduced much further. In practice $G^{(2)}(2N\pi)$ can be evaluated numerically by a second order central difference.

### 3.3. General error bounds

The numerical integration (32) has three error sources: the discretization error of the Gauss quadrature; the error from the tail approximation; and the error propagated from the error of the forward integration.

The error bound for the Gaussian quadrature, $\delta_G$, can be estimated from (20), (21) and (24)

$$\delta_G < 10^{-19} \sum_{k=0}^{2N-1} \Delta_k^{15} \sum_{j=1}^{n_k} C_{k,j} < 10^{-19} C_{\max} \sum_{k=0}^{2N-1} \Delta_k^{14} \sum_{j=1}^{n_k} \Delta_k = 10^{-19} \pi C_{\max} \sum_{k=0}^{2N-1} \Delta_k^{14}, \quad (33)$$

where $C_{k,j} = |g^{(14)}(\eta_{k,j})|$ is a constant in the sub-division $(a_{k,j}, b_{k,j})$, $a_{k,j} < \eta_{k,j} < b_{k,j}$, and $C_{\max} = \max(C_{k,j})$. By adding all the absolute errors from the subdivisions and taking the maximum $C_{k,j}$, (33) is a very conservative estimate.

The error from the tail approximation $\delta_T = |\varepsilon_T|$ is obtained from (31). For the propagation error, we estimate its bound $\delta_f$ as follows. Let $\varepsilon_R(t)$ be the error from numerical integration (14) for the real part of severity CF, and $\varepsilon_I(t)$ be the error from (15) for the imaginary part of severity CF. Then the error in df calculated using (32), due to errors $\varepsilon_R(t)$ and $\varepsilon_I(t)$, can be estimated from (16) as

$$\varepsilon_H(z) = \frac{2}{\pi} \int_0^\infty [e^{\lambda(\varphi_R(t)+\varepsilon_R(t)-1)} \cos(\lambda(\varphi_I(t)+\varepsilon_I(t))) - e^{\lambda(\varphi_R(t)-1)} \cos(\lambda\varphi_I(t))] \frac{\sin(tz)}{t} dt$$
$$\approx \lambda \frac{2}{\pi} \int_0^\infty e^{\lambda(\varphi_R(t)-1)} [\varepsilon_R(t)\cos(\lambda\varphi_I(t)) - \varepsilon_I(t)\sin(\lambda\varphi_I(t))] \frac{\sin(tz)}{t} dt, \quad (34)$$

where $\varphi_R(t) = \text{Re}[\varphi(t)]$ and $\varphi_I(t) = \text{Im}[\varphi(t)]$. The "$\approx$" sign in (34) is due to ignoring the higher order terms. In general, because $\varepsilon_R(t)$ and $\varepsilon_I(t)$ are unknown and random, (34) can not be computed precisely. However, we can evaluate (34) for two limiting cases at $z \to 0$ and $z \to \infty$. Let $\delta_C = \max(|\varepsilon_R(t)|, |\varepsilon_I(t)|)$ be the common error bound (e.g. specified by the user as the absolute error tolerance for the forward integration). Changing variable from $t$ to $y = tz$ and noting that: $\varphi_R(y/z) \to 0$, $\varphi_I(y/z) \to 0$ as $z \to 0$ and $\int_0^\infty \sin(y)/y\, dy = \pi/2$, (34) at $z = 0$ can be evaluated as

$$|\varepsilon_H(0)| = \lambda e^{-\lambda} |\varepsilon_R(\infty)| \le \lambda e^{-\lambda} \delta_C < \lambda \delta_C. \quad (35)$$

Similarly, as $z \to \infty$, $\varphi_R(y/z) \to 1$, $\varphi_I(y/z) \to 0$ and (34) becomes

$$|\varepsilon_H(\infty)| = \lambda |\varepsilon_R(0)| \le \lambda \delta_C, \quad (36)$$

The inequalities (35) and (36) show that the propagation error is proportional to the forward integration error bound. At the extreme case of $\lambda = 10^6$, a single precision can still be readily achieved if the forward integration has a double precision. As discussed in Section 3.1, the forward integration (14) and (15) can routinely achieve double precision accuracy through the use of *IMSL* functions based on Gauss-Kronrod quadrature. At any other point $0 < z < \infty$, it can be reasonably assumed that $\delta_f = \lambda \delta_C$ gives a conservative estimate for the error bound due to propagation.

The total absolute error in calculating $H(z)$ via (32) is bounded by the sum of $\delta_G$, $\delta_f$ and $\delta_T$, see formulas (33), (35) and (31):

$$10^{-19} \pi C_{\max} \sum_{k=0}^{2N-1} \Delta_k^{14} + \lambda \delta_C + \left| \sum_{k=1}^\infty (-1)^k G^{(2k)}(2N\pi) \right|. \quad (37)$$

It is worth pointing out that, for very large $\lambda$, the propagation error is likely the largest among the three error sources. In practice, (37) is not convenient to use because high order derivatives are involved, which is typical for analytical error bounds. As discussed in previous sections, an established and satisfactory practice is to use finer grids to estimate the error of the coarse grids.

## 4. Results for heavy tailed compound distribution

In this section we first validate the accuracy of our DNI algorithm for simple severity distributions with heavy tails, lognormal and GPD, without compounding. Then we present results for the 0.999 quantiles and CVaR of compound distributions, in comparison with those obtained by FFT and MC. For FFT we use *R* programming language (version 2.6.2) with the *actuar*-package freely available through CRAN. We attempt to provide relatively high precision results as benchmarks so that different methods (future and present) for computing compound distributions can be easily validated. This work is partly motivated by the lack of such comprehensive data in public domains.

The "best" estimates by each of three methods (DNI, FFT, MC) are presented in comparison. Here, the word "best" requires the following clarifications:

1. For DNI algorithm we start with a relatively coarse grid and short truncation length, and keep halving the grid size and doubling the truncation length, until the relative difference in 0.999 quantile is less than 0.01%. For this reason we show 5 digits for all results from all the three methods, unless stated otherwise. For all the following DNI calculations, we use the tail approximation (28), i.e. without the correction term $G^{(2)}(2N\pi)$;
2. For MC we do the maximum number of simulations, under the constraints of not exceeding $10^8$ or 24 hours, whichever takes less time. This means that for small to medium frequencies we perform $10^8$ simulations, but for large frequencies we have to do less number of simulations in order to get results within 24 hours;
3. For FFT we use the finest resolution $N_{FFT} = 2^{22}$ allowed on our PC when using *R*. The computation failed to yield any results for any higher value of $N_{FFT}$. The choice of suitable bandwidth $h$ for FFT requires careful attention. There is no analytically tractable formula available for the optimal gird size $h$ for minimizing both the discretization error and aliasing error. Embrechts and Frei (2008) suggested to successively reduce $h$ and compute the compound distribution with the rounding method until the improvement is smaller than some threshold. It is not clear from what level one should start this procedure. $h$ needs to be small in order to have a small discretization error, however, $h$ has to be large enough so that the truncation length $hN_{FFT} > \hat{Q}_q$, where $\hat{Q}_q$ is the quantile to be computed, which is not known a *prior*. As pointed out by Schaller and Temnov (2008), simply increasing $h$ for a given $N_{FFT}$ is the least efficient way to reduce the aliasing error and it is at the cost of increasing the discretization error. The exponential windowing based on an optimal choice of tilting parameter from a balance equation, as described by Schaller and Temnov (2008), offers an attractive alternative. In this study we only use standard FFT calculations without tilting. Also we use $h = \tilde{h}$ and $h = 2\tilde{h}$, where $\tilde{h} \equiv v \times 10^k$, and chose the smallest integer $1 \leq v \leq 9$ and $k$ ($k$ can be negative for $\tilde{h} < 1$) at which we get the quantile estimate $\hat{Q}_q$ satisfying the condition $N\tilde{h} > 2\hat{Q}_q$. This is to make sure the truncated length is at least twice as large as the desired quantile value and the choice of $h$ is systematic and consistent. In general, the magnitude of $\hat{Q}_q$ is unknown *a priori* and some iteration is required, which effectively reduces the actual speed advantage of FFT. Unless otherwise stated, we will show FFT results with grid spacing $h = \tilde{h}$ and $h = 2\tilde{h}$.

### *4.1. Validation of accuracy – no compounding*

Often in the literature the accuracy of the CF calculation is validated by applying the method to the cases where the transform is known in closed form. Here, we prefer to validate both steps (forward transform (2) and its inversion) of our DNI algorithm with the actual heavy tailed distributions. That is, we numerically integrate (2) and then invert the resulting CF, i.e. use (6) with $\chi(t) = \varphi(t)$ to get the df for the severity distribution (without compounding). Then, the latter is compared with the exact values. Once the accuracy is validated by this checking, we can assume that the CF evaluation for the compound distribution is accurate, since it only involves simple algebraic calculations to obtain a compound CF from a severity CF.

As described in Section 3.2.1, the user can specify the initial number of sub-divisions $n_0$, which effectively sets an initial grid size $\Delta_0 = \pi/n_0$. Another user specified controlling parameter is $N$, the number of full cycles, so the truncated length is $2\pi N$ at which we apply the tail integration approximation $G(2\pi N)$ given by (28).

In the first testing case, we compute the df for *Lognormal*(0,2), given double precision value $z = Q_{0.999} = 483.216412...$, for which the exact df value is simply 0.999. Table 1 shows results of various inputs, along with the FFT and MC estimates. The MC estimate is based on $N_{MC} = 10^8$ simulations and the FFT estimate is based on $N_{FFT} = 2^{22}$ discrete points (the largest size we were able to run *R* on our computer). The *discretize*() routine in *R actuar*-package was utilized to obtain the probability mass function (pmf) of the random variable through discretization of the df, before calling the FFT routine. We have tried different grid sizes with $h = 0.001, 0.01, 0.1, 0.2$ for the FFT calculation and found the optimal grid size for this case is $h = 0.1$. Both the MC and FFT results for the df have to be interpolated between discrete points. Results in Table 1 show that proposed DNI is the most accurate and the fastest among all three methods. A mere 2 seconds DNI computation gives higher accuracy than MC's 180 seconds and FFT's 14 seconds. All the calculations were carried out on a PC with a *Dual-Core AMD Opteron Processor 8220 SE*, 2.80 GHz, 3.75Gb RAM.

In the second testing case, we compute the df for *GPD*(1,1) at $z = Q_{0.999} = 999$, i.e. where the exact df value is 0.999. For the shape parameter $\xi = 1$, this is a very heavy tailed distribution for which mean and all higher moments do not exist. Table 2 shows DNI results along with the FFT and MC estimates. In this case we found $h = 0.5$ is better than $h = 0.001, 0.01, 0.1, 1.0$ in the FFT calculation. This illustrates one of the drawbacks of FFT – *a priori* we do not know the optimal gird size $h$ to use with a given $N_{FFT}$, especially if we do not known the correct answer. Once again, the comparison in Table 2 shows that DNI algorithm is the fastest and most accurate. The MC result showed a marked deterioration in accuracy for the GPD case, obviously due to the unbounded variance of this heavy tailed distribution.

It is interesting to demonstrate the effectiveness of the one-point tail approximation applied to heavy tailed distributions by doing the same calculations for *GPD*(1,1) *without* adding the tail approximation $G(2\pi N)$. We found the df error is $5.1 \times 10^{-4}$ for input $n_0 = 2, N = 100$, instead of $4.6 \times 10^{-9}$ as shown in Table 2, and the error is $7.1 \times 10^{-6}$ for input $n_0 = 2, N = 800$, instead of $1.6 \times 10^{-12}$. From another side, if the tail is ignored, the truncated domain needs to be extended to $N = 10^4$ in order to reduce the error to $5.1 \times 10^{-9}$, which is the same magnitude as with the tail approximation at $N = 100$. The CPU time for the integration extended to $N = 10^4$ is 121 seconds, which is 100 times longer than for $N = 100$ (see Table 2).

### *4.2. 0.999 quantiles for compound distributions*

Here, we attempt to obtain accurate estimates for the 0.999 quantile, $\hat{Q}_{0.999}$ of the Poisson-lognormal, Poisson-GPD and NegBinomial-lognormal compound distributions with heavy tails and loss frequencies ranging from 0.1 to one million. The DNI algorithm computes df, $H(z)$, for any given level $z$ by (6), one point at a time. In this regard, it is not an ideal setup for computing quantiles given df values. MC and FFT have the advantage in that they both obtain the whole distribution in a single run. With DNI we have to resort to an iterative procedure to inverse (6), requiring evaluation of (6) many times depending on the search algorithm employed and the initial guess. Here we employed a standard bisection algorithm until a convergence criterion $(\hat{H}(\hat{Q}_q) - q)/q < \varepsilon$ is satisfied. For $q = 0.999$, we set $\varepsilon = 10^{-12}$.

Table 3a shows the "best" estimates for *Poisson*($\lambda$)-*Lognormal*(0, 2) cases by DNI, MC and FFT. The case with no compounding (referred to as single severity), where "exact" (double precision) value is known, is also included. For this case we show 6 digits due to the high accuracy achieved by DNI. The relative error (using all digits) in comparison with the "exact" value ($Q_{0.999} = 483.216412...$) is $8.4 \times 10^{-8}$ for DNI, $1.3 \times 10^{-3}$ for MC and $1.4 \times 10^{-4}$ for FFT.

Table 3b shows the convergence of DNI results (six digits), starting from a coarse grid ($n_0 = 1$) and a short truncation length ($N = 50$). As the grid spacing is halved and the truncation length doubled successively, the relative change in estimated 0.999 quantile is reduced to less than 0.01%, in most cases this accuracy is confirmed after only the first refinement step. In other words, the relative error for the coarse grid in most cases is estimated to be less than 0.01%. In fact, the actual error for the coarse grid in the no-compound case (where there is an exact value to compare) is 0.0048%, which better than the "best" estimates from both MC and FFT. Interestingly, the relative change in predicted $\hat{Q}_{0.999}$ due to change from a coarse grid ($n_0 = 1, N = 50$) to the next finer grid ($n_0 = 2, N = 100$) is 0.0043% for the same case, which is very close to the actual error of 0.0048%.

In comparison, if we reduce $N_{FFT}$ from $2^{22}$ to $2^{21}$ and keep the same truncation length (this effectively doubles the grid spacing from $h = 0.0003$ to $h = 0.0006$), the FFT estimated $\hat{Q}_{0.999}$ will change by 0.014%. This larger relative difference is virtually the same as the corresponding actual error of $1.4 \times 10^{-4}$ in the estimated $\hat{Q}_{0.999}$. This nice agreement between relative change of value due to successive grid refinement and the actual relative error once again gives us confidence in estimating numerical error by the relative difference between successive grid refinements.

The agreement between the three methods is very good for all frequencies except for $\lambda = 10^6$. For all the cases with $\lambda \leq 10^5$, the maximum relative difference between the three methods is below 1%. At $\lambda = 10^6$ the difference between DNI and MC is still very small, about 0.1%, but the FFT value differs by more than 3%. It is worth pointing out that at this high frequency level, the FFT showed rather high sensitivity to grid size. For example, if we halve the gird size while keeping the same truncation length by letting $N_{FFT} = 2^{21}$, the estimated $\hat{Q}_{0.999}$ value changes by 3.7%, highlighting the problem that the optimal grid size for FFT is unknown. Every difference between DNI and MC is within the MC standard errors (given in brackets).

For $\lambda \leq 10^4$, the FFT with the grid spacing $h = 2\tilde{h}$ gave results closer to those of DNI than with the grid spacing $h = \tilde{h}$. The maximum relative difference between the two methods is less than 0.06%. However, for $\lambda > 10^4$, FFT with spacing $h = \tilde{h}$ gave results closer to DNI. We have a similar situation if we compare FFT with MC. FFT with grid spacing $h = 2\tilde{h}$ yields closer agreement with MC than with grid spacing $h = \tilde{h}$ for $\lambda \leq 10^4$, but for $\lambda > 10^4$ the grid spacing $h = \tilde{h}$ gave results in better agreement with MC. This highlights the problematic aspect of choosing the optimal gird spacing for FFT as there is no general rule to apply.

To obtain the results in Table 3a, the CPU time for MC ranges from a few minutes to more than 24 hours depending on $\lambda$. For FFT, each calculation for a given bandwidth $h$ took about 15 seconds. For DNI with the coarse grid ($n_0 = 1, N = 50$) and a crude initial guess, the CPU time on average is about 25 seconds. With a finer grid, we took the results of the coarse grid as the initial guess. As the coarse grid results are already very accurate, the additional number of iterations for the fine grids is small. In any case the coarse grid results were already sufficiently accurate in comparison with those shown in Table 3a, as demonstrated by the convergence shown in Table 3b. So, despite the iterative procedure required to calculate the inverse df using DNI, the DNI is much faster than MC for all frequencies, and it remains competitive with FFT. Note that FFT in general requires a few ad-hoc trials on the optimal bandwidth choice and its accuracy is relatively poor at a very high frequency, while the DNI can have relatively good accuracy with rather coarse grids at all the frequencies.

Table 4 shows results for *Poisson*($\lambda$)-*GPD*(1,1) cases by DNI, MC and FFT. With the shape parameter $\xi = 1$, the *GPD* has infinite mean and all higher moments. For the case with no compounding (single severity), the exact value $Q_{0.999} = 999$ is compared with numerical estimates. The relative error in comparison with the exact value is $4.3 \times 10^{-8}$ for DNI, $5.1 \times 10^{-4}$ for MC and $2.3 \times 10^{-4}$ for FFT. The convergence achieved for DNI is similar to the case of *Poisson*($\lambda$)-*Lognormal*(0,2) as shown in Table 3b.

The agreement between DNI and FFT with grid spacing $2\tilde{h}$ is very good for all values of $\lambda$, with the maximum relative error of less than 0.6% occurring at $\lambda = 10^6$, while MC has larger standard errors than for *Poisson*($\lambda$)-*Lognormal*(0, 2) cases.

For the FFT calculations, $N_{FFT}$ can be reduced to $2^{18}$ to obtain the same accuracy and higher speed for frequencies up to $10^3$, provided the grid size $h$ is increased at the same time by 16 times. In general, however, the true 0.999 quantile is unknown *a priori*, and the optimal combination of $N_{FFT}$ and $h$, even if we know the discretization error bound ($\varepsilon < \lambda h$), is not given by an analytically tractable formula. In other words, only if we are confident that $h \times N_{FFT}$ is sufficiently large, we can reduce $N_{FFT}$ and increase $h$ (keeping the error bound $\varepsilon < \lambda h$ in mind) to get higher speed for the same accuracy.

We also performed DNI computation for a very heavy tailed compound distribution *Poisson*($\lambda$)-*GPD*(1.5,1). The results for this case show that the following very simple scaling gives an excellent approximation

$$Q_{0.999}(\lambda) \approx Q_{0.999}(1)\lambda^\xi, \tag{38}$$

where $Q_{0.999}(1)$ is the 0.999 quantile for $\lambda = 1$. This is consistent with the closed form approximation for the heavy-tailed Poisson-GPD distribution (see e.g Böcker and Klüppelberg (2005), Embrechts, Klüppelberg and Mikosch (1997))

$$Q_q(\lambda) \sim \frac{\beta}{\xi}\left(\frac{\lambda}{1-q}\right)^\xi, \quad q \to 1, \tag{39}$$

where $q$ is a high quantile level. Figure 6 shows a comparison between DNI results and formula (38) for the case of *Poisson*($\lambda$)-*GPD*(1.5,1) on a log-scale plot. The maximum relative difference between the two is less than 0.3% and on average the relative error is about 0.05%, even less than between different numerical methods.

Table 5a shows the "best" 0.999 quantile estimates for *NegBinomial*(0.1, $m$)-*Lognormal*(0, 2) cases by DNI, MC and FFT. In these cases $p = 0.1$ and the frequency *K* has mean$(K) = m(1-p)/p = 9m$ and var$(K) = mean/p = 10 \times mean$. This variance is 10 times larger than in the case of *Poisson* for the same mean. This larger variance causes larger standard errors in the MC estimates. At high frequencies the agreement between DNI and FFT is closer than between DNI and MC or between FFT and MC, provided FFT uses grid spacing $2\tilde{h}$. Table 5b shows the convergence of the DNI calculations (6 digits). Again, we observe excellent convergence behaviour.

Approximation (38) or (39) also holds when the frequency distribution is negative binomial, in which case the parameter $\lambda$ is replaced by $m(1-p)/p$, the mean of negative binomial distribution. We do not show results for *NegBinomial*( $0.1, m$ )-*GPD* cases for lack of interest – they form a straight line on the log-log scale plot, similar to *Poisson*( $\lambda$ )-*GPD* cases.

### *4.3. Conditional Value at Risk*

Consider a rv Z with pdf $h(\cdot)$ and df $H(\cdot)$ in model (1). Then, the conditional value at risk (CVaR) at confidence level $q$ is defined as

$$CVaR_q[Z] = E[Z \mid Z \geq H^{-1}(q)] = \frac{E[Z]}{1-q} - \frac{1}{1-q}\int_0^{H^{-1}(q)} zh(z)dz, \qquad (40)$$

where $H^{-1}(.)$ is the quantile function, or the inverse function of $H(z)$, and $E[Z]$ is the mean of Z. Note that for model (1), the last equality in (40) holds only if $q \geq \Pr(Z=0)$, which is of primary interest in risk management, otherwise one has to account for discontinuity of the distribution function at $Z=0$. The CVaR is regarded as a better risk measure than the traditional value at risk (VaR), the latter is defined as the quantile $Q_q = H^{-1}(q)$ given a probability level $q$. The expectation $E[Z]$ is calculated as

$$E[Z] \equiv \mu_Z = E[K]E[X], \qquad (41)$$

where $E[X] = E[X_i]$, $i \geq 1$. For example, if $K$ is distributed from *Poisson*($\lambda$) and $X_i$ are from *Lognormal*($\mu, \sigma$), then $\mu_Z = \lambda \exp(\mu + \sigma^2/2)$. To compute (40), we have to calculate the quantile $Q_q$ first and then the expectation

$$\int_0^{Q_q} zh(z)dz = \frac{2}{\pi}\int_0^{Q_q} z\int_0^\infty \text{Re}[\chi(t)]\cos(tz)dtdz = \frac{2Q_q}{\pi}\int_0^\infty \text{Re}[\chi(x/Q_q)]\left[\frac{\sin x}{x} - \frac{1-\cos x}{x^2}\right]dx. \qquad (42)$$

In deriving (42), we use $h(z)$ given by expression (5) and change variable $x = tQ_q$. The computation procedure for $Q_q$ has already been described in previous sections. Recognizing that the term involving $\sin x / x$ corresponds to $H(Q_q) = q$, (40) becomes

$$CVaR_q[Z] = \frac{1}{1-q}\left[\mu_Z - qQ_q + \frac{2Q_q}{\pi}\int_0^\infty \text{Re}[\chi(x/Q_q)]\frac{1-\cos x}{x^2}dx\right]. \qquad (43)$$

The same DNI scheme as used for the computation of $H(Q_q)$ can be used for (43). Note for any given $q$, the evaluation of the quantile $Q_q$ is part of the calculation for CVaR.

Table 6 shows CVaR results for *Poisson*( $\lambda$ )-*Lognormal*(0,2) at $q = 0.999$. In the calculations for DNI we use $Q_{0.999}$ as estimated by DNI in Table 3a. The agreement among the three methods is good, with the exception of FFT failing completely for $\lambda \geq 10^5$. For FFT, we used formula similar to (40), i.e. the CVaR is computed by the difference between the mean and the expectation below the threshold. Otherwise, FFT estimation of CVaR appears to be very inaccurate if we directly take the expectation above the threshold. Still, at high frequencies FFT completely failed and using different grid size $h$ did not resolve this problem. Therefore, considering both accuracy and speed, DNI provides a very attractive alternative to MC and FFT for computing CVaR, especially at high frequencies

Table 7 shows CVaR results for *NegBinomial*($0.1, m$)-*Lognormal*($0, 2$) cases. Once again, the FFT failed completely for high frequencies ($m \geq 10^4$, mean($K$) $\geq 9 \times 10^4$). For both *Poisson*($\lambda$)-*Lognormal*($0,2$) and *NegBinomial*($0.1, m$)-*Lognormal*($0, 2$) cases, the DNI computation of the CVaR also showed excellent convergence behaviour with different grid sizes and truncation lengths, similar to the quantile calculations in Tables 3b and 5b.

The CPU time taken by DNI to compute (43), in addition to the quantile calculation, is also given in Tables 6 and 7. In practice, sometimes we may wish to compute the expected exceedance above a loss amount $L$, i.e. (43) with $H^{-1}(q)$ replaced by $L$. Then there is no need to compute the quantile, making DNI a very attractive candidate for such a task.

One of the commonly recognised drawbacks of FFT is the so-called *aliasing* error associated with the finite truncation length. Embrechts and Frei (2008) and more recently Schaller and Temnov (2008) demonstrated the use of the exponential "windowing" or tilting technique to reduce aliasing errors in FFT convolution of heavy tailed distributions. It was shown that the application of exponential tilting can significantly reduce the aliasing errors, provided a proper value of tilting parameter is used. Schaller and Temnov (2008) found the optimal value for the tilting parameter by solving a balance equation. It is worth pointing out that at least for frequencies up to $10^3$, application of tilting with the tilting parameter estimated from the balance equation of Schaller and Temnov (2008) can lead to the same accuracy with reduced $N_{FFT}$ and thus much higher speed for FFT. However, in our current case FFT still failed completely for high frequencies ($\lambda \geq 10^5$ or $m \geq 10^4$) with tilting, regardless the value of the tilting parameter, though such high frequency may not be relevant to OpRisk practice. In general, when the discretization error is already large, tilting makes little difference for the overall accuracy because tilting reduces the aliasing errors only.

## 5. Conclusions

We have implemented an efficient numerical convolution DNI algorithm for computing high quantiles and conditional Value at Risk of compound distributions with heavy tails and a wide range of frequencies. The efficiency (accuracy and speed) of this algorithm mainly comes from an innovative tail integration approximation and an adaptive grid spacing strategy. The usual truncation error associated with finite length of truncated integration domain can be reduced dramatically by employing the tail integration approximation, at virtually no extra computing cost, so a higher accuracy is achieved with a shorter truncation length.

For quantile and CVaR calculations of the compound distributions with moderate to high frequencies and heavy tails this DNI algorithm is not only faster than Monte Carlo, as expected, but also competitive with the very fast FFT, with the same or better accuracy. The FFT may completely fail to compute CVaR at very high frequencies and the MC takes too long to get accurate results, while DNI can get accurate results for a very wide range of frequencies. For low to medium frequencies up to $10^3$, the application of exponential tilting with the optimal value for the tilting parameter from the balance equation of Schaller and Temnov (2008) can significantly reduce the aliasing errors of FFT, enabling a much faster calculation.

Although we have focused on some specific distributions, e.g. Poisson and negative binomial for frequency and lognormal and Pareto for severity relevant to OpRisk, the method can be used in general. The scope of the paper is restricted to independent severities and frequencies. It is more challenging to compute the compound distribution accurately and efficiently when there is dependence between the random variables. The extension of the present DNI algorithm to include dependence between risks is a subject of future research.

## Acknowledgement

We would like to thank David Gates, Mark Westcott and three anonymous referees for many constructive comments which have led to significant improvements in the manuscript. In particular, we are grateful to one of the referees for suggesting the use of formula (31).

## Appendix

Consider a non-negative random variable $Z$ with density $h(z)$, $z \geq 0$. Then its CF is

$$\chi(t) = \int_{-\infty}^{\infty} h(z) e^{itz} dz = \text{Re}[\chi(t)] + i \, \text{Im}[\chi(t)], \tag{A1}$$

where

$$\text{Re}[\chi(t)] = \int_0^{\infty} h(z) \cos(tz) dz, \quad \text{Im}[\chi(t)] = \int_0^{\infty} h(z) \sin(tz) dz. \tag{A2}$$

Now, define a function $\tilde{h}(z)$ such that $\tilde{h}(z) = h(z)$ if $z \geq 0$ and $\tilde{h}(z) = h(-z)$ if $z < 0$. The CF for this extended function is (using symmetry property)

$$\tilde{\chi}(t) = \int_{-\infty}^{\infty} \tilde{h}(z) e^{itz} dz = 2 \int_0^{\infty} h(z) \cos(tz) dz = 2 \text{Re}[\chi(t)], \quad \tilde{\chi}(t) = \tilde{\chi}(-t). \tag{A3}$$

Thus, the density $h(z) = \tilde{h}(z), z \geq 0$ can be retrieved as

$$h(z) = \frac{1}{2\pi} \int_{-\infty}^{\infty} \tilde{\chi}(t) e^{-itz} dt = \frac{1}{\pi} \int_0^{\infty} \tilde{\chi}(t) \cos(tz) dt = \frac{2}{\pi} \int_0^{\infty} \text{Re}[\chi(t)] \cos(tz) dt \tag{A4}$$

and the distribution can be calculated as

$$H(z) = \int_0^z h(y) dy = \int_0^z \frac{2}{\pi} dy \int_0^{\infty} \text{Re}[\chi(t)] \cos(ty) dt = \frac{2}{\pi} \int_0^{\infty} \text{Re}[\chi(t)] \frac{\sin(tz)}{t} dt. \tag{A5}$$

Changing variable $x = t \times z$, (A5) can be rewritten as

$$H(z) = \frac{2}{\pi} \int_0^{\infty} \text{Re}[\chi(x/z)] \frac{\sin(x)}{x} dx. \tag{A6}$$

Thus

$$H(z \to 0) = \frac{2}{\pi} \text{Re}[\chi(\infty)] \int_0^\infty \frac{\sin(x)}{x} dx = \text{Re}[\chi(\infty)]. \qquad (A7)$$

Note that (A7) leads to correct limit for compound distribution, $H(0) = \Pr(K = 0)$, because the severity CF $\varphi(\infty) \to 0$ (in the case of continuous severity df), also see (4). For example, $H(0) = e^{-\lambda}$ for compound $Poisson(\lambda)$, and $H(0) = p^m$ for compound $NegBinomial(p,m)$, see (11) and (12) respectively.

**Table 1.** Comparison of the estimated $Lognormal(0,2)$ df, $\hat{F}$, at the 0.999 quantile, calculated via DNI, FFT and MC methods. The error is $\varepsilon = (\hat{F} - 0.999)/0.999$.

| Method | Input | Error $\varepsilon$ | CPU (seconds) |
|---|---|---|---|
| DNI | $n_0 = 2$, $N = 100$ | $7.3 \times 10^{-9}$ | 2 |
| | $n_0 = 4$, $N = 100$ | $3.7 \times 10^{-9}$ | 2 |
| | $n_0 = 8$, $N = 200$ | $3.6 \times 10^{-10}$ | 6 |
| | $n_0 = 16$, $N = 400$ | $2.6 \times 10^{-11}$ | 12 |
| MC | $N_{MC} = 10^8$ | $2.3 \times 10^{-6}$ | 181 |
| FFT | $h = 0.1$, $N_{FFT} = 2^{22}$ | $8.9 \times 10^{-8}$ | 15 |

**Table 2.** Comparison of the estimated $GPD(1,1)$ df, $\hat{F}$, at 0.999 quantile, calculated via DNI, FFT and MC methods. The calculated error is $\varepsilon = (\hat{F} - 0.999)/0.999$.

| Method | Input | Error $\varepsilon$ | CPU (seconds) |
|---|---|---|---|
| DNI | $n_0 = 2$, $N = 100$ | $4.6 \times 10^{-9}$ | 1 |
| | $n_0 = 2$, $N = 200$ | $4.7 \times 10^{-10}$ | 2 |
| | $n_0 = 4$, $N = 400$ | $4.0 \times 10^{-11}$ | 5 |
| | $n_0 = 4$, $N = 800$ | $1.9 \times 10^{-12}$ | 11 |
| MC | $N_{MC} = 10^8$ | $1.0 \times 10^{-3}$ | 122 |
| FFT | $h = 0.5$, $N_{FFT} = 2^{22}$ | $1.1 \times 10^{-8}$ | 14 |

**Table 3a.** "Best" 0.999 quantile estimates $\hat{Q}_{0.999}$ for $Poisson(\lambda)$-$Lognormal(0,2)$, calculated by DNI, FFT and MC methods. Standard errors of MC are given in brackets. SS is the case of single severity (without compounding).

| $\lambda$ | DNI $\hat{Q}_{0.999}$ | MC $\hat{Q}_{0.999}$ (Std Err.) | $N_{MC}$ | FFT ($N_{FFT} = 2^{22}$) $\hat{Q}_{0.999}$ | $h$ |
|---|---|---|---|---|---|
| SS | 483.216 | 482.576 (1.81) | $10^8$ | 483.285 | 0.0003 |
| | | | | 483.285 | 0.0006 |
| 0.1 | 105.38 | 105.46 (0.498) | $10^8$ | 105.29 | 0.00006 |
| | | | | 105.37 | 0.00012 |
| 1 | 490.55 | 490.91 (1.84) | $10^8$ | 490.00 | 0.0003 |
| | | | | 490.54 | 0.0006 |
| 10 | 1779.2 | 1780.7 (5.31) | $10^8$ | 1771.7 | 0.0009 |
| | | | | 1778.5 | 0.0018 |
| $10^2$ | 5853.1 | 5861.7 (14.2) | $10^8$ | 5809.6 | 0.003 |
| | | | | 5849.8 | 0.006 |
| $10^3$ | 21149 | 21150 (32.3) | $10^8$ | 21139 | 0.02 |
| | | | | 21150 | 0.04 |
| $10^4$ | 1.0835E5 | 1.0832E5 (182) | $1.5 \times 10^7$ | 1.0830E5 | 0.06 |
| | | | | 1.0834E5 | 0.12 |
| $10^5$ | 8.2235E5 | 8.2233E5 (1340) | $1.5 \times 10^6$ | 8.2141E5 | 0.4 |
| | | | | 8.1946E5 | 0.8 |
| $10^6$ | 7.5974E6 | 7.5909E6 (5660) | $1.5 \times 10^5$ | 7.3479E6 | 4 |
| | | | | 7.0762E6 | 8 |

**Table 3b. Convergence in the quantile estimates $\hat{Q}_{0.999}$ by DNI for *Poisson*($\lambda$)-*Lognormal*(0,2). SS is the case of single severity (without compounding).**

| $\lambda$ | $\hat{Q}_{0.999}$ ($n_0=1, N=50$) | $\hat{Q}_{0.999}$ ($n_0=2, N=100$) | $\hat{Q}_{0.999}$ ($n_0=4, N=200$) | $\hat{Q}_{0.999}$ ($n_0=8, N=400$) |
|---|---|---|---|---|
| SS | 483.193 | 483.214 | 483.216 | 483.216 |
| 0.1 | 105.349 | 105.361 | 105.383 | 105.383 |
| 1 | 490.527 | 490.527 | 490.549 | 490.549 |
| 10 | 1779.15 | 1779.16 | 1779.16 | 1779.16 |
| $10^2$ | 5853.06 | 5853.06 | 5853.06 | 5853.06 |
| $10^3$ | 21149.4 | 21149.4 | 21149.4 | 21149.4 |
| $10^4$ | 1.08354E5 | 1.08354E5 | 1.08354E5 | 1.08354E5 |
| $10^5$ | 8.22350E5 | 8.22350E5 | 8.22350E5 | 8.22350E5 |
| $10^6$ | 7.58869E6 | 7.59749E6 | 7.59745E6 | 7.59745E6 |

**Table 4. The 0.999 quantile estimates $\hat{Q}_{0.999}$ for *Poisson*($\lambda$)-*GPD*(1,1), calculated by DNI, FFT and MC methods. SS is the case of single severity (without compounding).**

| $\lambda$ | DNI $\hat{Q}_{0.999}$ | MC $\hat{Q}_{0.999}$ (StdErr.) | $N_{MC}$ | FFT $\hat{Q}_{0.999}$ | h |
|---|---|---|---|---|---|
| SS | 998.999 | 998.488 (6.33) | $10^8$ | 999.238 | 0.0005 |
|  |  |  |  | 999.238 | 0.001 |
| 0.1 | 99.353 | 99.4292636 (0.625) | $10^8$ | 99.271 | 0.00005 |
|  |  |  |  | 99.348 | 0.0001 |
| 1 | 1004.9 | 1004.17 (6.39) | $10^8$ | 1003.5 | 0.0005 |
|  |  |  |  | 1004.7 | 0.001 |
| 10 | 10081 | 9998.69 (61.6) | $10^8$ | 10062 | 0.005 |
|  |  |  |  | 10078 | 0.01 |
| $10^2$ | 1.0105E5 | 100724 (634) | $10^8$ | 1.0081E5 | 0.05 |
|  |  |  |  | 1.0101E5 | 0.1 |
| $10^3$ | 1.0128E6 | 1011206 (6323) | $10^8$ | 1.0099E6 | 0.5 |
|  |  |  |  | 1.0122E6 | 1 |
| $10^4$ | 1.0151E7 | 1.0125E7 (1.15E5) | $1.5 \times 10^7$ | 1.0114E7 | 5 |
|  |  |  |  | 1.0137E7 | 10 |
| $10^5$ | 1.0174E8 | 9.9222E8 (4.55E6) | $1.5 \times 10^6$ | 1.0119E8 | 50 |
|  |  |  |  | 1.0141E8 | 100 |
| $10^6$ | 1.0197E9 | 1.0849E9 (1.76E8) | $1.5 \times 10^5$ | 1.0119E9 | 500 |
|  |  |  |  | 1.0141E9 | 1000 |

**Table 5a. The 0.999 quantile estimates $\hat{Q}_{0.999}$ for *NegBinomial*(0.1,m)-*Lognormal*(0,2), calculated by DNI, FFT and MC methods.**

| m | mean(N) | DNI $\hat{Q}_{0.999}$ | MC $\hat{Q}_{0.999}$ (StdErr) | $N_{MC}$ | FFT $\hat{Q}_{0.999}$ | h |
|---|---|---|---|---|---|---|
| 1 | 10 | 1763.8 | 1761.3 (5.16) | 1E8 | 1755.1 | 0.001 |
|  |  |  |  |  | 1763.1 | 0.002 |
| 10 | $9 \times 10$ | 5631.6 | 5634.39 (13.6) | 1E8 | 5594.3 | 0.003 |
|  |  |  |  |  | 5628.9 | 0.006 |
| $10^2$ | $9 \times 10^2$ | 19961 | 20033.57 (34.1) | 1E8 | 19816 | 0.01 |
|  |  |  |  |  | 19953 | 0.02 |
| $10^3$ | $9 \times 10^3$ | 99935 | 99846 (151) | 2E7 | 99833 | 0.05 |
|  |  |  |  |  | 99927 | 0.1 |
| $10^4$ | $9 \times 10^4$ | 7.4664E5 | 7.4610E5 (992) | 2E6 | 7.4580E5 | 0.4 |
|  |  |  |  |  | 7.4403E5 | 0.8 |
| $10^5$ | $9 \times 10^5$ | 6.8576E6 | 6.8551E6 (4370) | 2E5 | 6.6324E6 | 4 |
|  |  |  |  |  | 6.3873E6 | 8 |

**Table 5b. Convergence in the quantile estimates by DNI for *NegBinomial*(0.1,*m*)-*Lognormal*(0,2).**

| m | $\hat{Q}_{0.999}$ ($n_0 = 1$, $N = 50$) | $\hat{Q}_{0.999}$ ($n_0 = 2$, $N = 100$) | $\hat{Q}_{0.999}$ ($n_0 = 4$, $N = 200$) | $\hat{Q}_{0.999}$ ($n_0 = 8$, $N = 400$) |
|---|---|---|---|---|
| 1 | 1763.84 | 1763.84 | 1763.84 | 1763.84 |
| 10 | 5631.63 | 5631.63 | 5631.63 | 5631.63 |
| $10^2$ | 19961.2 | 19961.2 | 19961.2 | 19961.2 |
| $10^3$ | 99935.0 | 99935.0 | 99935.0 | 99935.0 |
| $10^4$ | 746638 | 746638 | 746638 | 746638 |
| $10^5$ | 6.84063E6 | 6.85759E6 | 6.85760E6 | 6.85760E6 |

**Table 6. Comparison of CVaR results for *Poisson*($\lambda$)-*Lognormal*(0,2) calculated by DNI, FFT and MC methods. For DNI, $\hat{Q}_{0.999}$ is taken from Table 3a, and the CPU is the time required to compute (43) in addition to computing $\hat{Q}_{0.999}$. The MC standard errors are given in brackets next to the MC estimates.**

| $\lambda$ | L | DNI CVaR | DNI CPU | MC CVaR | FFT CVaR | h |
|---|---|---|---|---|---|---|
| 0.1 | 105.38 | 275.58 | 16 sec | 275.59 (1.55) | 275.49 | 0.00006 |
|  |  |  |  |  | 275.51 | 0.00012 |
| 1 | 490.55 | 1026.1 | 12 sec | 1024.4 (3.62) | 1026.7 | 0.0003 |
|  |  |  |  |  | 1025.9 | 0.0006 |
| 10 | 1779.2 | 3241.8 | 12 sec | 3249.8 (9.32) | 3241.1 | 0.0009 |
|  |  |  |  |  | 3242.4 | 0.0018 |
| $10^2$ | 5853.1 | 9470.7 | 6 sec | 9453.3 (20.0) | 9460.5 | 0.003 |
|  |  |  |  |  | 9469.9 | 0.006 |
| $10^3$ | 21149 | 29421 | 2 sec | 29290 (41.1) | 29431 | 0.02 |
|  |  |  |  |  | 29508 | 0.04 |
| $10^4$ | 1.0835E5 | 1.2605E5 | 2 sec | 1.2576E5 (218) | 1.2808E5 | 0.06 |
|  |  |  |  |  | 1.3560E5 | 0.012 |
| $10^5$ | 8.2235E5 | 8.5761E5 | 4 sec | 8.5573E5 (1127) | 1.7578E6 | 0.4 |
|  |  |  |  |  | 3.6517E6 | 0.8 |
| $10^6$ | 7.5974E6 | 7.6599E6 | 11 sec | 7.6531E6 (5927) | 2.4992E8 | 4 |
|  |  |  |  |  | 5.2139E8 | 8 |

**Table 7. Comparison of CVaR results for *NegBinomial*(0.1,*m*)-*Lgnormal*(0,2) calculated by DNI, FFT and MC methods. For DNI, $\hat{Q}_{0.999}$ is taken from Table 5a, and the CPU is the time required to compute (43) in addition to computing $\hat{Q}_{0.999}$. The MC standard errors are given in brackets next to the MC estimates.**

| m | L | DNI CVaR | DNI CPU | MC CVaR (StdErr) | FFT CVaR | h |
|---|---|---|---|---|---|---|
| 1 | 1763.8 | 3159.6 | 12 sec | 3164.1 (8.78) | 3159.9 | 0.001 |
|  |  |  |  |  | 3161.9 | 0.002 |
| 10 | 5631.6 | 9102.4 | 12 sec | 9095.2 (19.8) | 9093.0 | 0.003 |
|  |  |  |  |  | 9101.9 | 0.006 |
| $10^2$ | 19961 | 27918 | 2 sec | 27774 (40.5) | 27829 | 0.01 |
|  |  |  |  |  | 27928 | 0.02 |
| $10^3$ | 99935 | 1.1697E5 | 2 sec | 1.1654E5 (176) | 1.1801E5 | 0.05 |
|  |  |  |  |  | 1.2283E5 | 0.1 |
| $10^4$ | 7.4664E5 | 7.8047E5 | 3 sec | 7.7872E5 (958) | 1.5917E6 | 0.4 |
|  |  |  |  |  | 3.2980E6 | 0.8 |
| $10^5$ | 6.8576E6 | 6.9167E6 | 9 sec | 6.9022E6 (3674) | 2.2496E8 | 4 |
|  |  |  |  |  | 4.6926E8 | 8 |

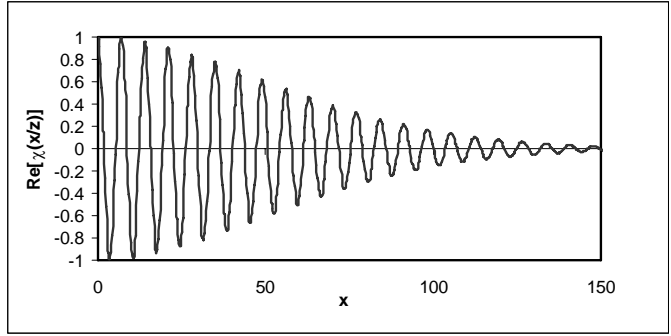

**Figure 1.** $\operatorname{Re}[\chi(x/z)]$ **for** *Lognormal(0,2)-Poisson($10^5$)*. $z = 8.22 \times 10^5 \approx Q_{0.999}$.

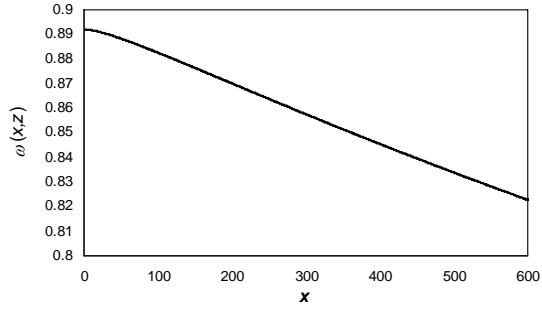

**Figure 2. Frequency ratio** $\omega(x,z)$ **for** *Lognormal(0,2)-Poisson($10^5$)*. $z = 8.22 \times 10^5 \approx Q_{0.999}$.

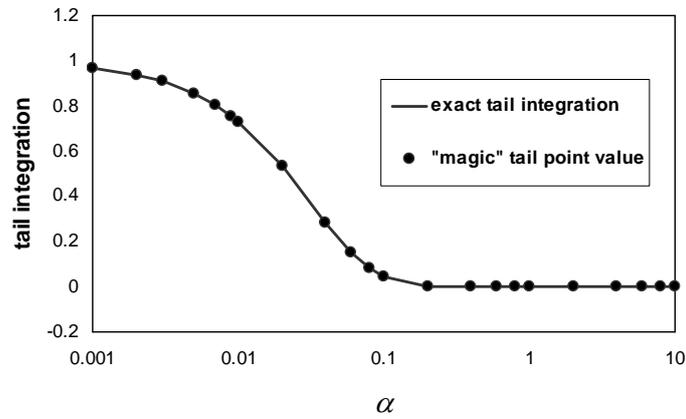

**Figure 3a. Comparison between the exact tail integration** $\int_{2\pi N}^{\infty} G(x)\sin(x)dx$ **and simple one-point approximation** $G(2N\pi)$ **from formula (28), when** $G(x) = e^{-\alpha x}$ **and** $N = 5$.

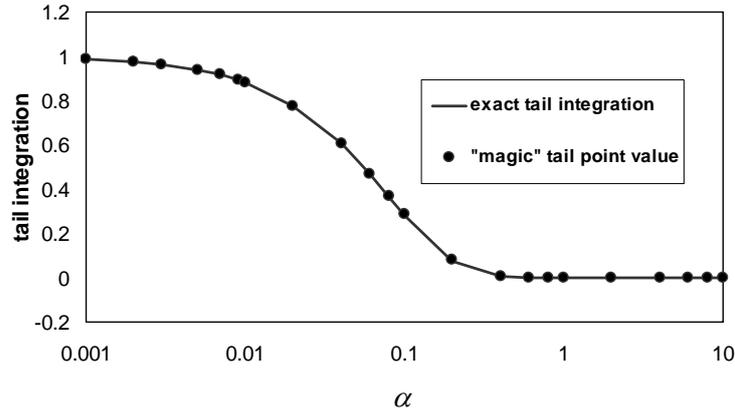

**Figure 3b.** Comparison between the exact tail integration $\int_{2\pi N}^{\infty} G(x)\sin(x)dx$ and simple one-point approximation $G(2N\pi)$ from formula (28), with $G(x) = e^{-\alpha x}$ and $N = 2$.

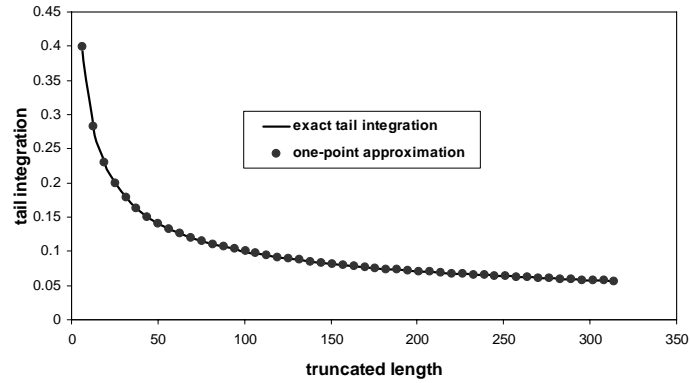

**Figure 4a.** Comparison between exact tail integration $\int_{2\pi N}^{\infty} G(x)\sin(x)dx$ and the simple one point approximation (28), $G(2N\pi)$, as functions of truncated length $l_T = 2N\pi, 2 \leq N \leq 50$, when $G(x) = 1/\sqrt{x}$.

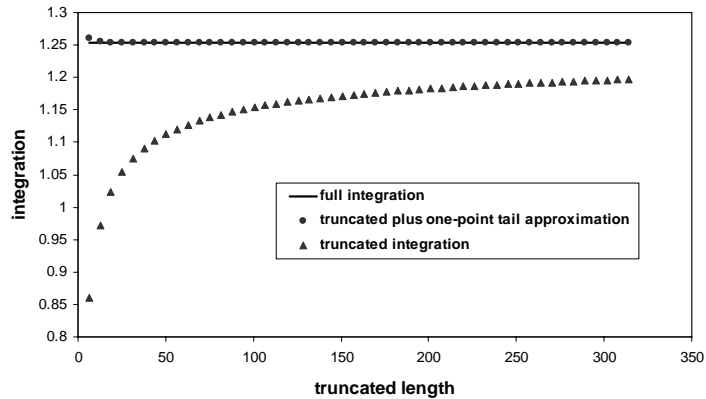

**Figure 4b.** Comparison between truncated integration $\widetilde{I}(2N\pi) = \int_0^{2\pi N} G(x)\sin(x)dx$ and the truncated integration plus the one-point approximation of tail integration, $\widetilde{I}(2N\pi) + G(2N\pi)$, as functions of the truncated length $l_T = 2N\pi, 2 \leq N \leq 50$, where $G(x) = 1/\sqrt{x}$. The solid line represents the exact value of the full integration without truncation error, $I_E = \widetilde{I}(\infty) = \sqrt{\pi/2}$.

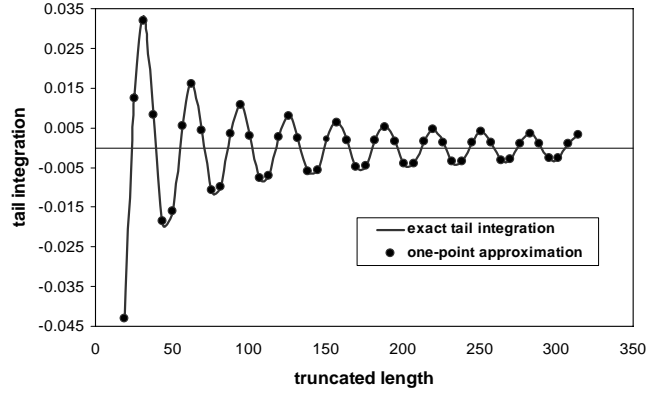

**Figure 5a.** Comparison between exact tail integration $\int_{2\pi N}^{\infty} G(x)\sin(x)dx$ and the simple one point approximation (28), $G(2N\pi)$, as a function of truncated length $l_T = 2N\pi, 3 \leq N \leq 50$. $G(x) = \cos(\alpha x)/x$.

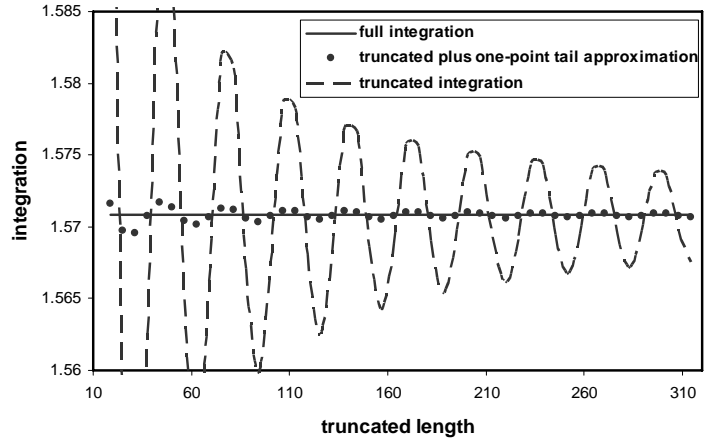

**Figure 5b.** Comparison between truncated integration $\widetilde{I}(2N\pi) = \int_0^{2\pi N} G(x)\sin(x)dx$ and the truncated integration plus the one-point approximation of tail integration, $\widetilde{I}(2N\pi) + G(2N\pi)$, as functions of the truncated length $l_T = 2N\pi, 3 \leq N \leq 50$, when $G(x) = \cos(\alpha x)/x$. The solid line represents the exact value of the full integration without truncation error, $I_E = \widetilde{I}(\infty) = \pi/2$.

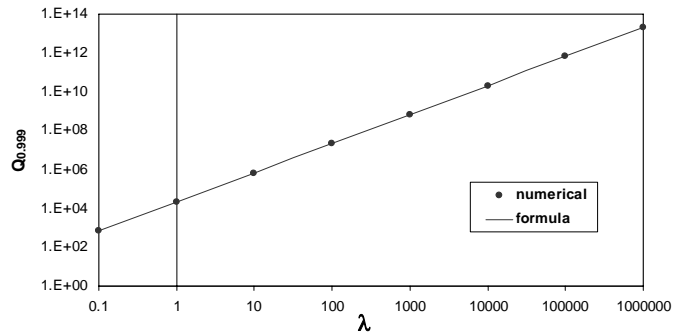

**Figure 6.** Comparison between numerical results and formula (38) for compound distribution *Poisson*($\lambda$)-*GPD*(1.5,1). Log scale is used for both axes.

33